\documentclass[aps,twocolumn,showpacs,prl,superscriptaddress]{revtex4-2}
\usepackage{mathtools}
\usepackage{bm}
\usepackage{dsfont,amsthm,amsbsy}
\usepackage{verbatim}
\usepackage{amssymb}
\usepackage{amsmath}
\usepackage{bbm}
\usepackage{graphicx}
\usepackage{epstopdf}
\usepackage{subfigure}
\usepackage{natbib}
\usepackage{epsfig}
\usepackage{amsfonts}
\usepackage{mathrsfs}
\usepackage{sidecap}
\usepackage{lipsum}
\usepackage[toc,page,title,titletoc,header]{appendix}
\usepackage[colorlinks,linkcolor=blue,citecolor=blue,anchorcolor=blue,urlcolor=blue]{hyperref}
\usepackage{hyperref}
\usepackage{resizegather}
\usepackage{tikz}
\usepackage{float}
\usepackage{mathbbol}
\usepackage[normalem]{ulem}
\usepackage{cancel}
\usepackage{upgreek}

\newcommand{\bl}{\begin{aligned}}
\newcommand{\el}{\end{aligned}}

\def\be{\begin{equation}}
\def\ee{\end{equation}}

\def\bi{\begin{itemize}}
\def\ei{\end{itemize}}
\def\bn{\begin{enumerate}}
\def\en{\end{enumerate}}
\def\bea{\begin{eqnarray}}
\def\eea{\end{eqnarray}}

\def\ba{\begin{array}}
\def\ea{\end{array}}
\def\bd{\begin{displaymath}}
\def\ed{\end{displaymath}}

\begin{document}
%=====================================================
\title
{Separation of the Kibble-Zurek Mechanism from Quantum Criticality}

%=====================================================

\author{R. Jafari}
\email[]{raadmehr.jafari@gmail.com}
\affiliation{Department of Physics, Institute for Advanced Studies in Basic Sciences (IASBS), Zanjan 45137-66731, Iran}
\affiliation{School of Quantum Physics and Matter, Institute for Research in Fundamental Sciences (IPM), Tehran 19538-33511, Iran}
%\affiliation{Physics Department and Research Center OPTIMAS, University of Kaiserslautern, 67663 Kaiserslautern, Germany}

\author{Alireza  Akbari}
\email[]{alireza@bimsa.cn}
\affiliation{Beijing Institute of Mathematical Sciences and Applications (BIMSA), Huairou District, Beijing 101408, China}
\affiliation{Max Planck Institute for the Chemical Physics of Solids, D-01187 Dresden, Germany}

\date{\today}

\begin{abstract}
When a system is swept through a quantum critical point (QCP), the Kibble–Zurek mechanism predicts that the average number of topological defects follows a universal power-law scaling with the ramp time scale. This scaling behavior is determined by the equilibrium critical exponents of the underlying phase transition. We show that the correspondence between Kibble–Zurek scaling and quantum criticality does not hold generally. In particular, the defect density can exhibit a suppression faster than the Kibble–Zurek prediction even when the quench crosses a critical point, while conventional Kibble–Zurek scaling may persist for quenches through a {\em non-critical} point. Our results, based on models representative of a broad class of quasi-one-dimensional Fermi systems, identify the dynamical conditions under which universal defect scaling emerges and clarify the relation between defect generation and equilibrium criticality.
\end{abstract}

\pacs{}
\maketitle
%-----------------------------------------------------------------------------
{\em Introduction} $-$
Understanding nonequilibrium quantum dynamics, particularly across quantum phase transitions (QPTs), is crucial both for fundamental physics and for emerging quantum technologies. This includes phenomena ranging from many-body correlations~\cite{Abanin2019,Nandkishore2015}, quantum matter~\cite{Goldman2014,Wilczek2012}, and quantum simulations~\cite{Prufer2018,Erne2018}, to practical applications in quantum technologies~\cite{Buluta2009,Georgescu2014,Albash2018}.
The central question in  nonequilibrium quantum physics is how to characterize and control quantum critical behaviors and dynamics of 
the QPTs. %quantum phase transitions. 
The critical point where a quantum phase transition takes place is typically associate with 
a vanishing energy gap~\cite{Sachdev2004}, and as the system crosses the critical point, excitations (defects) are unavoidably created.
%Thus, challenging the achievement of an adiabatic driving across quantum critical point (QCP), irrespective of the path chosen to acquire this crossing.
This fundamentally challenges the possibility of achieving perfectly adiabatic driving across a quantum critical point  QCP, irrespective of the chosen path.

The Kibble–Zurek mechanism (KZM) provides a widely used framework in nonequilibrium statistical physics for analyzing the dynamics of systems driven across a QCP.
The KZM, first proposed in the cosmological context~\cite{Kibble1976,Zurek1985}, 
has been studied in vastly different contexts encompassing holographic superconductors, junction arrays, ion crystals, 
classical and quantum spin systems, fermionic and bosonic atomic and helium superfluids~\cite{Yates1998,Lagun1997,Dziarmaga2008,Kolodrubetz2012,Jelic2011,Sonner2015,Sonner2015,Chesler2015,Silvi2016,Damski2010,
Liu2018,Dora2019,Polkovnikov2005,Antunes2006,Grandi2011,Kolodrubetz2012b,Chandran2012,Hu2015,Liu2015,
Logan2016,Subires2022,Hwang2015,Adolfo2018,Berdanier2017,Nikoghosyan2016,Grabarits2025,Subires2022,Grabarits2025b,Minchul2015,Suzuki2010,Das2012,Oshiyama2020,Deffner2017,Maegochi2022,Jamadagni2024,Ladewig2020,Das2021,Sinha2019,Dziarmaga2022,Zurek2005,Dziarmaga2005,Sen2008,Bermudez2009,Shreyoshi2009}, 
with direct experimental verification in a broad range of different physical platforms~\cite{Mark1994,Maniv2003,Corman2014,Navon2015,Zamora2020,Keesling2019,Zhang2017,Bando2020,Higuera2022}.

The key prediction of the KZM is that the average topological defect density, $n_d$, represents a universal power-law scaling with the ramp time scale $\tau$, i.e., $n_d=\tau^{-\beta}$. This universal behavior is entirely governed by the equilibrium 
critical exponents of the phase transition, $\beta=d\nu/(1+z\nu)$, where $\nu$ and $z$ are the correlation length and 
dynamic critical exponents, respectively, and $d$ is the dimensionality of the system. 
%
%{\color{blue}Although KZM is widely associated with QCPs, conflicting evidence, including anti-Kibble–Zurek behavior and KZ-like scaling at non-critical points, raises a fundamental puzzle: is criticality truly essential for KZM?} 
However, a conflicting observations have been reported in the study of ferroelectric phase transition \cite{Griffin2012} and open systems \cite{Nigmatullin2016,Dario2008,Dario2009,Anirban2016,Puebla2020,Sadeghizade2025,Balducci2023,Singh2021,Singh2023,Gao2017,Jafari2026a}. This is classified as anti-Kibble-Zurek (AKZ) behavior, which characterized by an increase in the number of excitations with the ramp time scale.

In this Letter we challenge the notion that quantum criticality and KZM are intrinsically linked. 
While this notion has been supported by numerous studies, all studies conducted thus far have only explored models 
in one- and two-dimensional systems \cite{Dziarmaga2005,Sen2008,Bando2020,Zurek2005,Bermudez2009,Shreyoshi2009,Higuera2022,Liou2018,Das2017,Arnold2022,Zeng2025,Yuan2025} where the quasi-particles that dictate the excitation probability are the same as those that become massless at the critical point.
However, by extending our analysis beyond this limited class of models, where the excitation probabilities are determined by quasi-particles that remain massive at the critical point, the anticipated connection between Kibble-Zurek scaling and criticality breaks down.
To illustrate this, we analyze three fully connected critical systems: the generalized compass model (GCM)~\cite{You2014,Jafari2011,Eriksson2009,Jafari2012,Nikolaev2018,Sun2009,Brzezicki2007,Nussinov2015}, 
the transverse-field Ising model with Dzyaloshinsky-Moriya (DM) interaction~\cite{Moriya1960,Dzyaloshinsky1958,Jafari2008,Antal1997,Wang2018,Haldar2020,Roy2019}, 
and the generalized XY (GXY) model~\cite{Titvinidze2003,Zvyagin2006,Divakaran2013,Jafari2017}.
\\

%
%%%%%%%%%%%%%%%%%%%%%%%%%%%%  Fig.1 %%%%%%%%%%%%%%%%%%%%%%%%%%%%%%%%
\begin{figure*}[t!]
\centerline{
\includegraphics[width=0.33\linewidth]{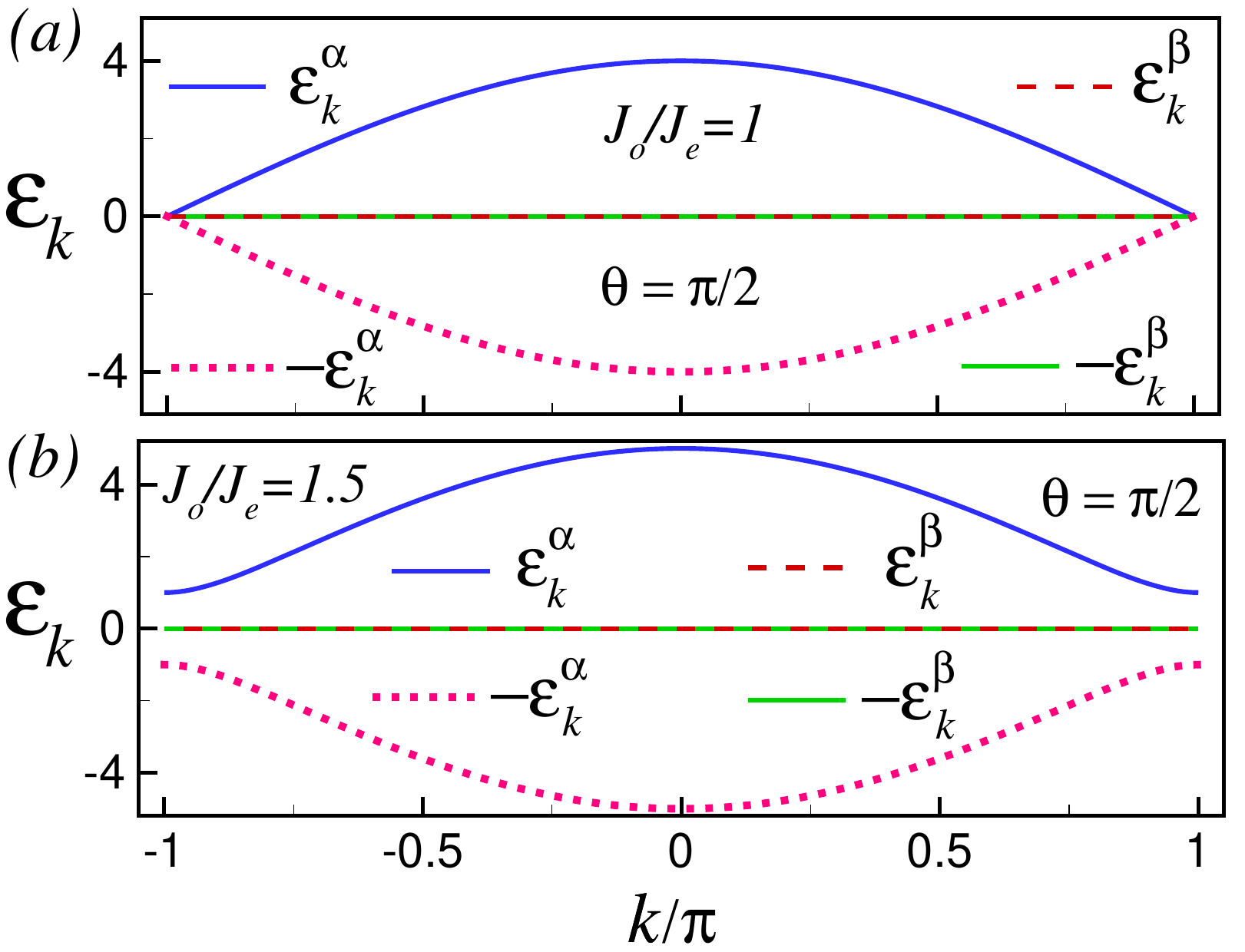}
\includegraphics[width=0.33\linewidth]{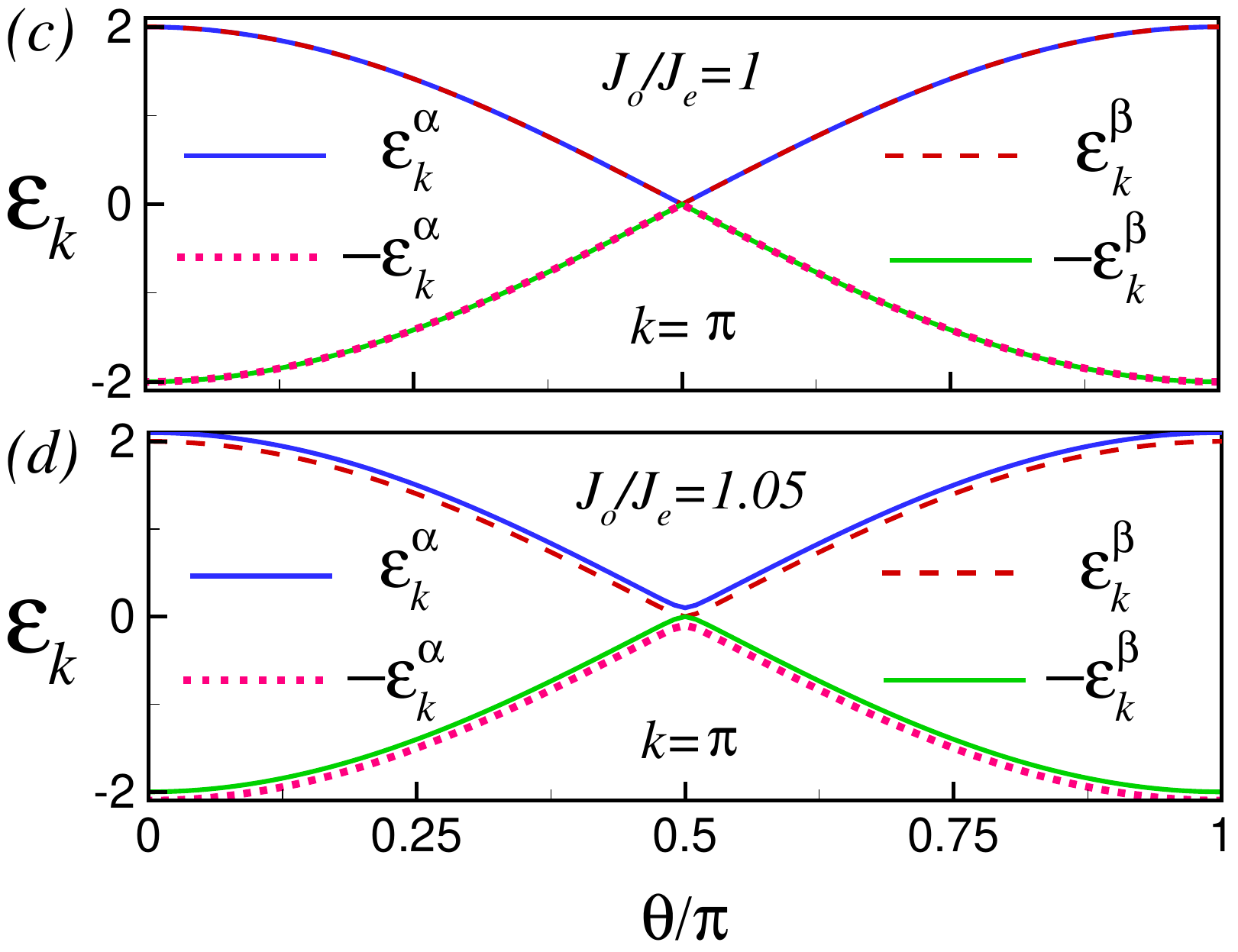}
\includegraphics[width=0.33\linewidth]{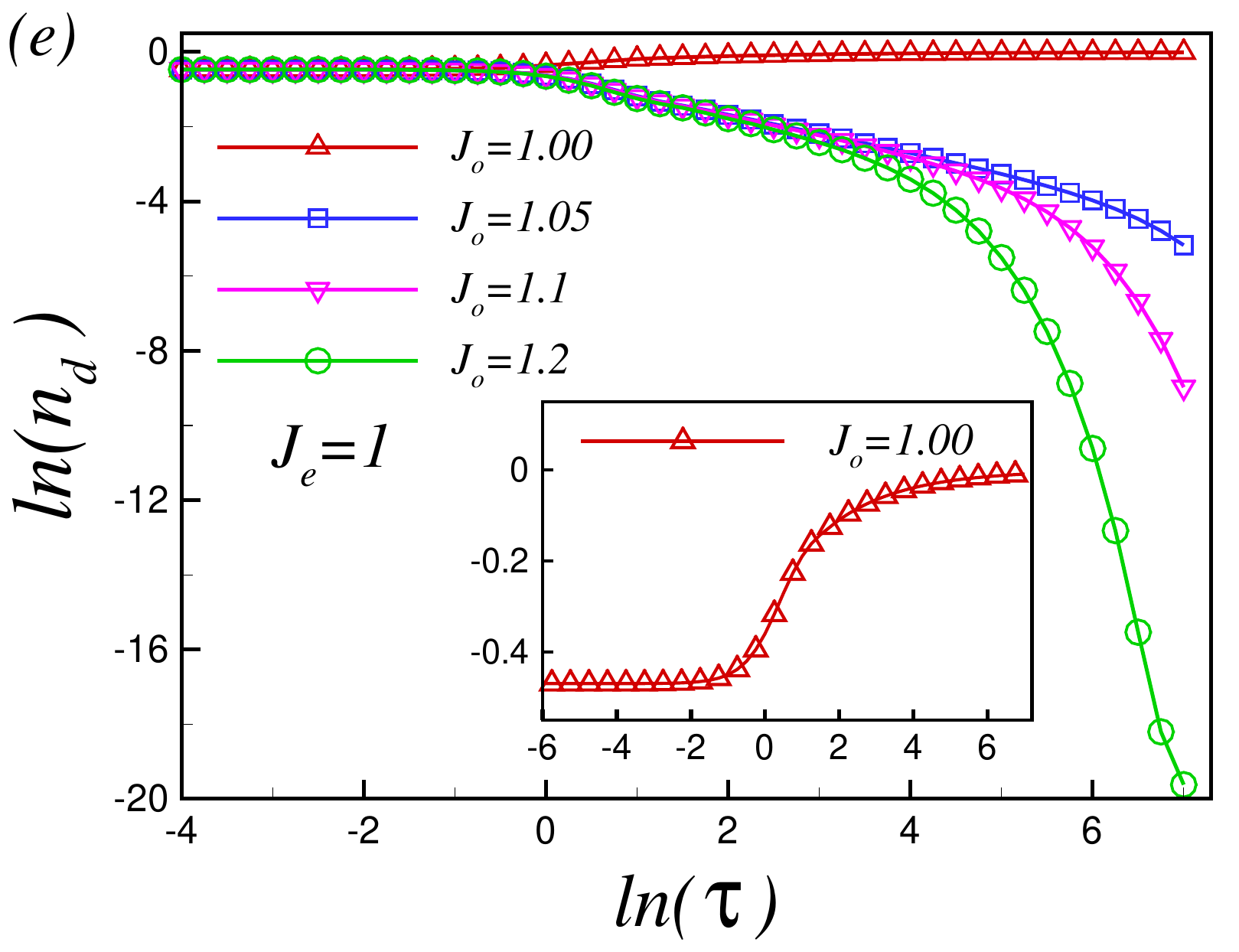}}
\caption{The quasi-particle spectrum $\pm\varepsilon_{k}^{\alpha,\beta}$ 
of the generalized compass model versus $k$ at the critical point $\theta_c=\pi/2$ at (a) the 
isotropic point (IP) $J_o=J_e=1$, and at (b) the anisotropic point $J_o=1.5$, $J_e=1$.
The quasi-particle spectrum versus $\theta$ at $k=\pm\pi$ at (c) the isotropic point (IP) $J_o=J_e=1$, 
and at (d) the anisotropic point $J_o=1.05$, $J_e=1$. (e) The density of excitations $n_d$ in 
the generalized compass model as a function of ramp time scale $\tau$ for quench from $\theta_i=0$ to 
$\theta_f=\pi$ for $J_e=1$ and different values of $J_o=1, 1.05, 1.1$ and $1.2$, for the system size $N=1024$. 
Inset represents the density of defects at the IP. }
\label{fig1}
\end{figure*}
%%%%%%%%%%%%%%%%%%%%%%%%%%%%%%%%%%%%%%%%%%%%%%%%%%%%%%%%%%%%%%%
%

A central finding of the present work is that,
%A key result of the present work is that,
 the dynamics of driven GCM across the critical point illustrate the circumstances under which the topological defects show suppression significantly faster than what KZM predicts. 
As our finding is confirmed in the transverse field Ising model with DM interaction (TFIMDM), however, the density of defects shows KZ scaling for a quench through the {\em non-critical} point in TFIMDM.
In light of this information and the results derived from the GXY model, we conclude that quantum criticality is neither a necessary nor a sufficient condition for KZ scaling and non-adiabatic dynamics. 
The pivotal aspect is that if the quasi-particle modes that govern the dynamics are massless, the driven is non-adiabatic.  
The KZM is only associated with QPT when these modes coincide with the quantum critical modes.

These general conditions shed new light on an important paradigm in nonequilibrium statistical physics. Moreover, as the adiabatic driving of quantum many-body systems is of interest to a wide variety of quantum technologies ranging from quantum simulation to adiabatic quantum computation, our findings can illuminate new perspectives on approaches which circumvent the KZ scaling law to achieve adiabatic dynamics~\cite{Doria2011,Adolfo2012,Campbell2015,Puebla2020b,Mathey2010}.
\\

{\em Generalized Compass Model} $-$ 
To lay the foundation, we consider the 1D spin-$1/2$ generalized compass Hamiltonian~\cite{You2014}
%
%%%%%%%%%%%%%%%%%%%%%%%%%%%%%%%%%%%  Eq.~ECH %%%%%%%%%%%%%%%%%%%%%%%%%%%%%%
\begin{equation}
\bl
\label{eq:GCH}
{\cal H}
\!=\!
\sum_{n=1}^{N}
\big[J_o\tilde{\sigma}_{2n-1}(\theta)\tilde{\sigma}_{2n}(\theta)
\!+\!
J_e\tilde{\sigma}_{2n}(-\theta)\tilde{\sigma}_{2n+1}(-\theta) \big],
\el
\end{equation}
%%%%%%%%%%%%%%%%%%%%%%%%%%%%%%%%%%%%%%%%%%%%%%%%%%%%%%%%%%%%%%%%%%%%%%%
%%%%%%%%%%%%%%%%%%%%%%
%
with $J_{e/o}$ exchange amplitudes on even/odd lattice bonds,
where the pseudo-spin operators $\tilde{\sigma}_n(\pm \theta)$ are
formed by linear combinations of the Pauli matrices $\sigma_{n}^{x}$ and $\sigma_{n}^{y}$:
$\tilde{\sigma}_n(\pm \theta) =\cos(\theta) \sigma_{n}^{x} \mp \sin(\theta) \sigma_{n}^{y}$. 
The Hamiltonian can be diagonalized exactly by mapping it onto a free fermion model,
%
%
%%%%%%%%%%%%%%%%%%%%%%%%%%%%%%%%%%%  Eq.~FFM %%%%%%%%%%%%%%%%%%%%%%%%%%%%%%
\begin{equation}
\bl
\label{eq:FFM}
{\cal H}=&
\sum_{n=1}^{N}
\Big[(J_{o} c^{\dagger}_{2n}c_{2n-1}+J_{e} c^{\dagger}_{2n+1}c_{2n})\\
%\no
&+(J_{o} e^{-i\theta} c^{\dagger}_{2n}c^{\dagger}_{2n-1}+J_{e} e^{i\theta}c^{\dagger}_{2n+1}c^{\dagger}_{2n})+\mbox{H.c.}\Big],
\el  
\end{equation}
%%%%%%%%%%%%%%%%%%%%%%%%%%%%%%%%%%%%%%%%%%%%%%%%%%%%%%%%%%%%%%%%%%%%%%%
using the Jordan-Wigner transformation~\cite{Jafari2025SM}. By partitioning the chain into bi-atomic
elementary cells and defining two independent fermions at each cell $n$~\cite{Perk1975,Titvinidze2003,Derzhko2009,Sun2009,Jafari2011,Jafari2017}, 
$c_{n}^{q}\equiv c_{2n-1}$ and $c_{n}^{p}\equiv c_{2n}$, performing the Fourier transformation and introducing the Nambu spinors
$C^{\dagger}= (c^{q\dagger}_{k},c^{q}_{-k},c^{p\dagger}_{k},c^{p}_{-k})$, the Hamiltonian in Eq.~(\ref{eq:FFM}) expressed as a sum
over decoupled mode Hamiltonians ${\cal H}= \sum_{k\ge0}C^{\dagger} {\cal H}_{k} C$, with
%
%%%%%%%%%%%%%%%%%%%%%%%%%%%%%%%%%%%  Eq.~FBH %%%%%%%%%%%%%%%%%%%%%%%%%%%%%%
\bea
\label{BdG}
{\cal H}_{k}=
\left(
  \begin{array}{cccc}
    0 & 0 & L_{k} & J_{k} \\
    0 & 0 & -J_{-k}^{\ast} & L_{-k}^{\ast} \\
    L_{k}^{\ast} & -J_{-k} & 0 & 0 \\
    J_{k}^{\ast} & L_{-k} & 0 & 0 \\
  \end{array}
\right),
\eea
%%%%%%%%%%%%%%%%%%%%%%%%%%%%%%%%%%%%%%%%%%%%%%%%%%%%%%%%%%%%%%%%%%%%%%%
%
where, $J_{k}=J_{o}e^{i \theta}- J_e e^{i (k-\theta)}$, and $L_{k}=J_{o}+ J_e e^{\it ik}$ 
and $k=(2m-1)\pi/N$, $m=1,\cdots,N/2$, given periodic boundary conditions.
%%
%%%%%%%%%%%%%%%%%%%%%%%%%%%%%  Fig.2 %%%%%%%%%%%%%%%%%%%%%%%%%%%%%%%%
%\begin{figure*}[t]
%\centerline{
%\includegraphics[width=0.33\linewidth]{Fig2a.pdf}
%\includegraphics[width=0.33\linewidth]{Fig2b.pdf}
%\includegraphics[width=0.33\linewidth]{Fig2a.pdf}}
%\caption{(Color online) BdG quasi-particle spectrum $\{\pm\varepsilon_{k}^{1,2}\}_{k=-\pi}^{\pi}$ for the ESSH model at
%(a) the isotropic point (IP) $w=\Delta=\tau=\Lambda=1$, and
%(b) at the anisotropic point $w=\Delta=2$, $\tau=\Lambda=1$.}
%\label{fig2}
%\end{figure*}
%%%%%%%%%%%%%%%%%%%%%%%%%%%%%%%%%%%%%%%%%%%%%%%%%%%%%%%%%%%%%%%%
%%
By diagonalizing ${\cal H}_{k}$ one obtains the quasi-particle Hamiltonian 
%
%%%%%%%%%%%%%%%%%%%%%%%%%%%%%%%%%%%  Eq.~FBH %%%%%%%%%%%%%%%%%%%%%%%%%%%%%%
\bea
\label{QPH}
{\cal H}_{k}=\sum_{k}\Big[\varepsilon^{{\alpha}}_{k}(\gamma_{k}^{{\alpha}\dag}\gamma_{k}^{{\alpha}}-1/2)
+\varepsilon^{{\beta}}_{k}(\gamma_{k}^{{\beta}\dag}\gamma_{k}^{{\beta}}-1/2)\Big],
\eea
%%%%%%%%%%%%%%%%%%%%%%%%%%%%%%%%%%%%%%%%%%%%%%%%%%%%%%%%%%%%%%%%%%%%%%%
%
with $\gamma_{k}^{{\alpha,\beta}\dag}$ and $\gamma_{k}^{{\alpha,\beta}}$ linear combinations of the elements in the Nambu spinor, 
and with corresponding energy bands $\varepsilon^{\alpha}_{k}=\sqrt{2(r_k+\sqrt{r_k^{2}-s_k}~)}$ and 
$\varepsilon^{\beta}_{k}=\sqrt{2(r_k-\sqrt{r_k^{2}-s_k}~)}$, where $r_k=|J_{k}|^{2}+|L_{k}|^{2}+|J_{-k}|^{2}+|L_{-k}|^{2}$ and
$s_k=4(L_{k}^{\ast}L_{-k}-J_{k}^{\ast}J_{-k})(L_{k}L_{-k}^{\ast}-J_{k}J_{-k}^{\ast})$. 
The ground state $|\Psi_0\rangle$ is obtained by filling up the negative-energy quasi-particle states,
$|\Psi_0\rangle = \prod_k |\Psi_{0,k} \rangle= \prod_k \gamma_k^{\alpha \dag} \gamma_k^{\beta \dag} |V\rangle_k$, where $|V\rangle_k$ is the Bogoliubov vacuum annihilated by the $\gamma_k$'s~\cite{Jafari2025SM}.
\\

It is straightforward to show that the band gap between $\varepsilon^{\beta}_{k}$ and $-\varepsilon^{\beta}_{k}$ vanishes for all momenta $k$ at 
$\theta_c=\pi/2$, for any ratio of $J_o/J_e$ [see Figs.~\ref{fig1}(a)-(d)]. 
Therefore, a QPT occurs at $\theta_c=\pi/2$ between two gapped spin phases, characterized by long-range spin correlations in the $x$ and $y$ directions, 
for any value of $J_o/J_e$~\cite{Nussinov2015}. 
In addition, at the isotropic point (IP) $J_o/J_e=1$, the band gap between $\varepsilon^{\alpha}_{k}$ and $-\varepsilon^{\alpha}_{k}$ closes at $\theta_c=\pi/2$ exclusively at the boundary modes $k=\pm\pi$,
as shown in Figs.~\ref{fig1}(a) and \ref{fig1}(c). 
Away from the isotropic point (IP), that is for $J_o/J_e \neq 1$, the band gap between $\varepsilon^{\beta}_{k}$ and $-\varepsilon^{\beta}_{k}$ remains zero for all momenta $k$, as shown in Figs.~\ref{fig1}(b) and \ref{fig1}(d). In contrast, a finite gap of magnitude $|J_e - J_o|$ opens at the Brillouin zone boundaries $k = \pm \pi$, separating the bands $\varepsilon^{\alpha}_{k}$ and $-\varepsilon^{\alpha}_{k}$.
Consequently, at the critical point $\theta_c=\pi/2$, where the model exhibits maximal frustration of interactions, the ground state exhibits a macroscopic degeneracy of $2^{N/2}$ when $J_e\neq J_o$, which increases to $2\times 2^{N/2}$ at the isotropic point $J_o/J_e=1$.
\\

 %
%%%%%%%%%%%%%%%%%%%%%%%%%%%%  Fig.2 %%%%%%%%%%%%%%%%%%%%%%%%%%%%%%%%
\begin{figure*}[]
\centerline{
\includegraphics[width=0.245\linewidth]{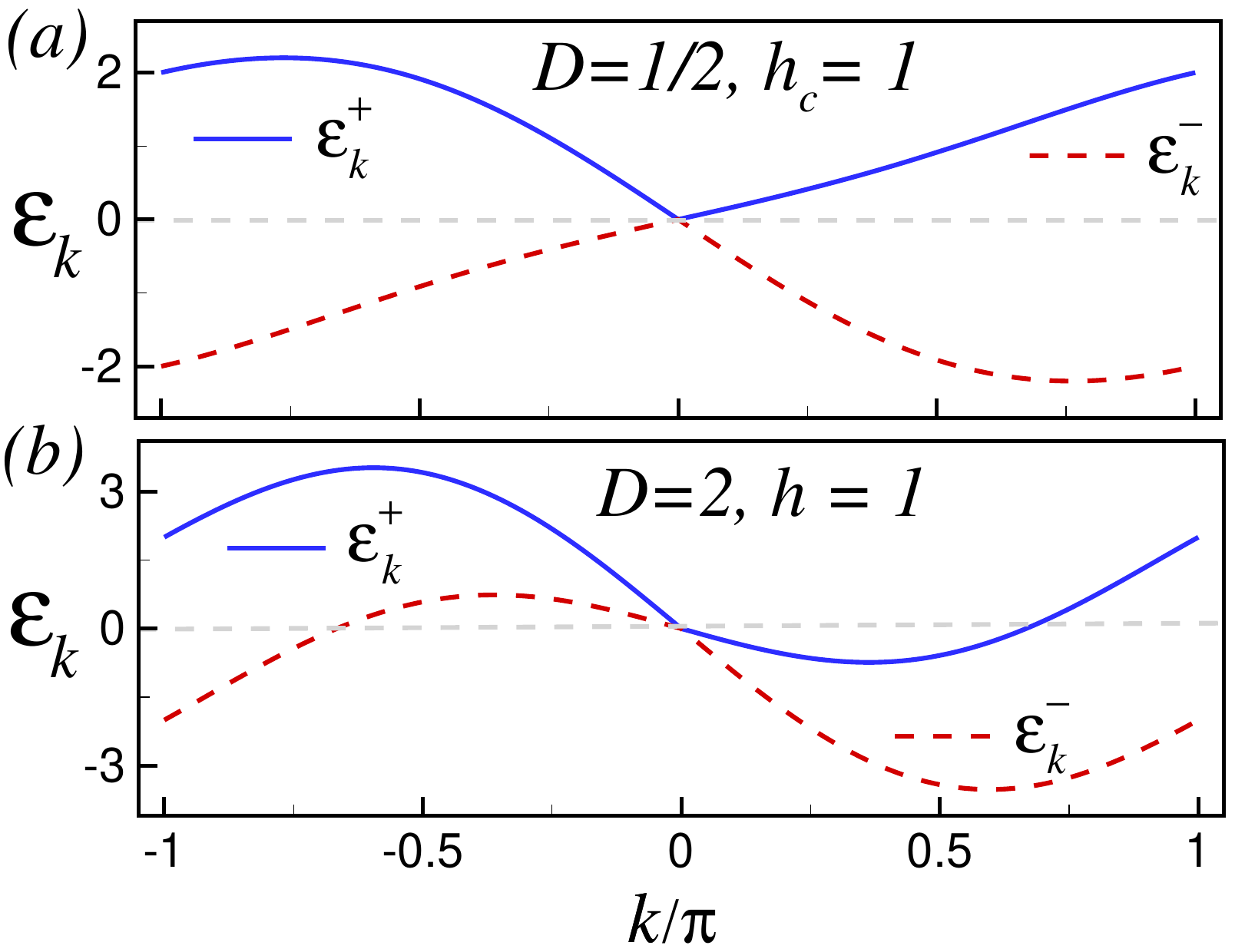}
\includegraphics[width=0.245\linewidth]{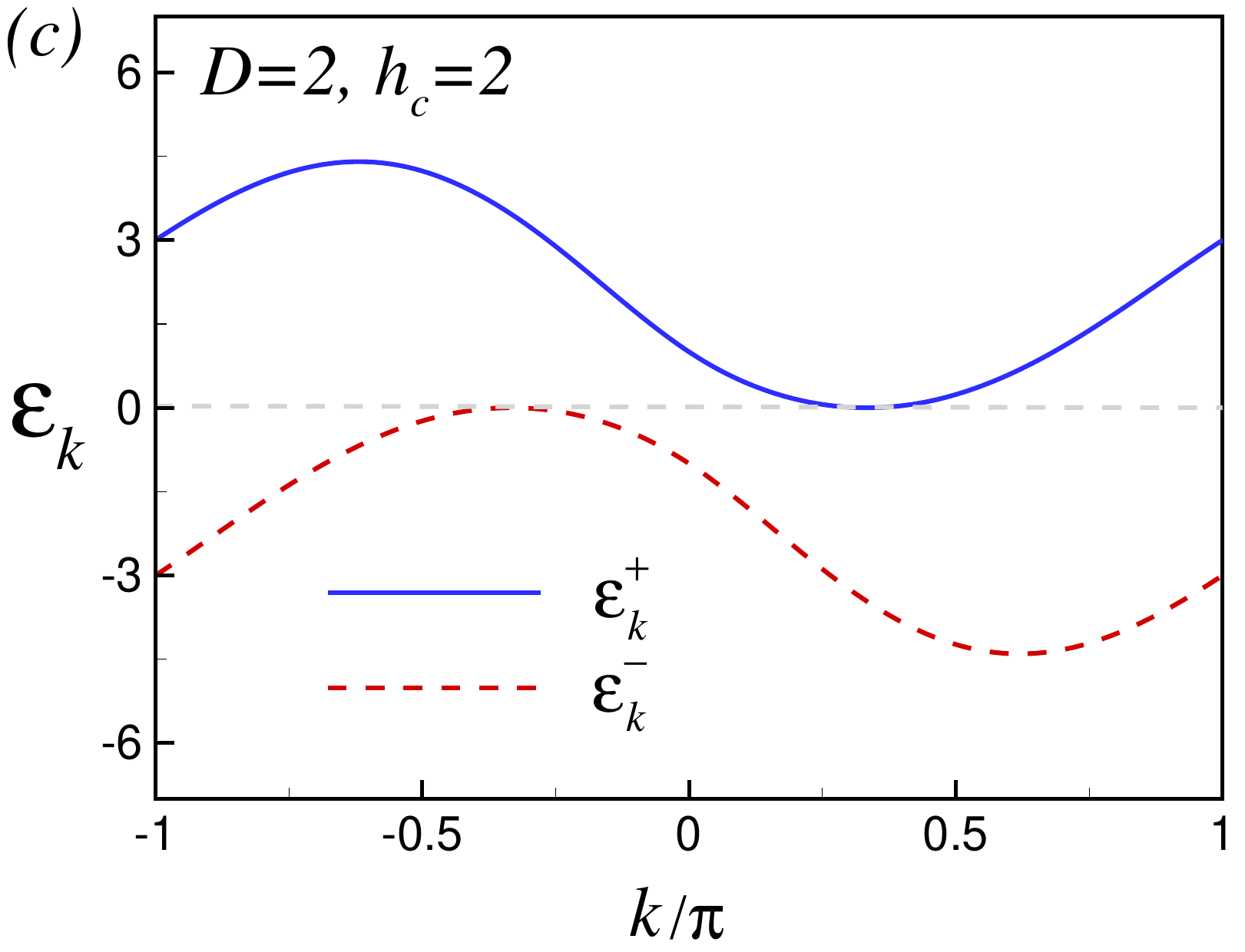}
\includegraphics[width=0.24\linewidth]{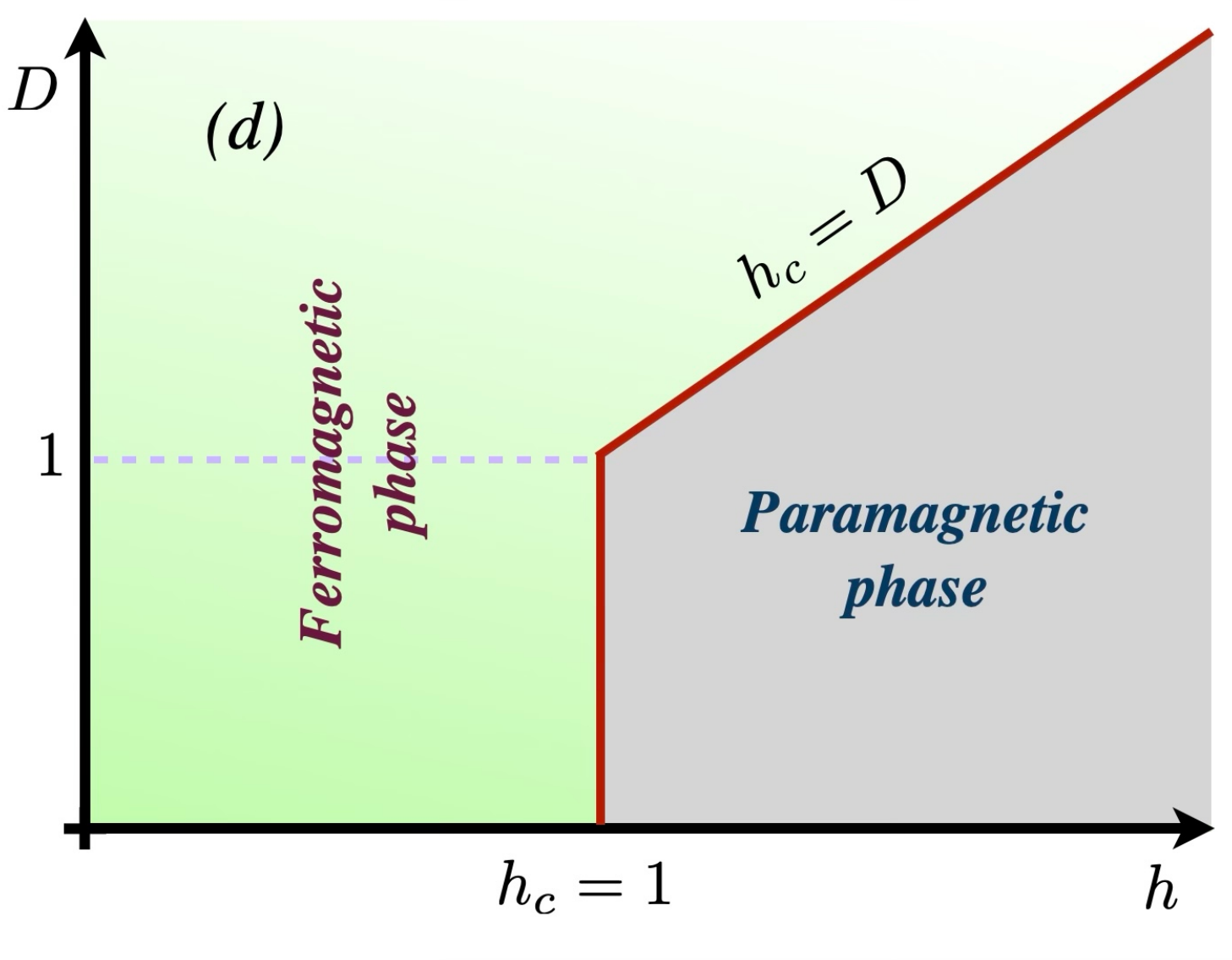}
\includegraphics[width=0.245\linewidth]{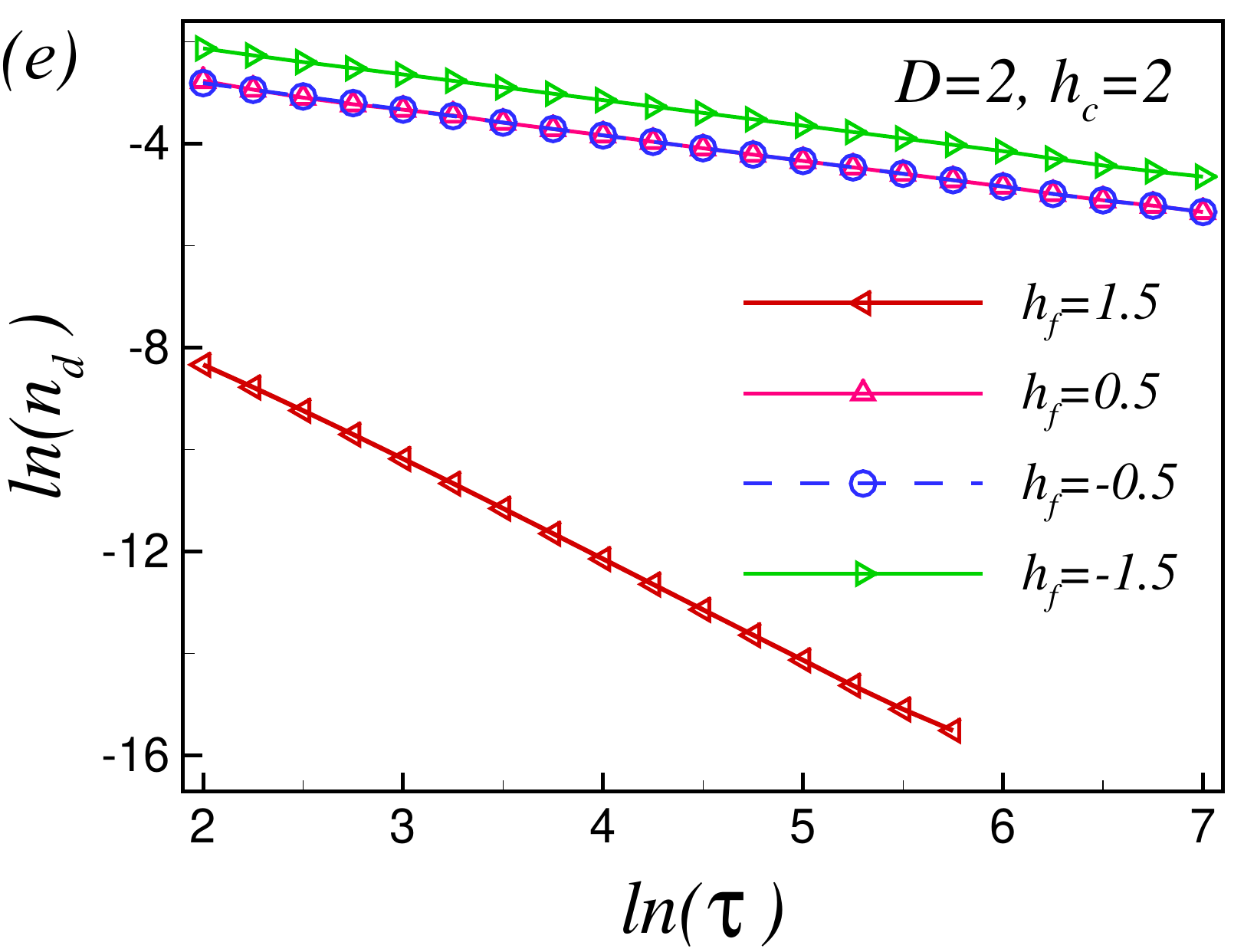}}
\caption{Quasi-particle spectrum $\pm \varepsilon_{k}^{\pm}$ versus momentum $k$: 
 (a) at the critical point $h_c=1$ for $D = 1/2$, (b) at $h=1$ and  $D = 2$, and (c) at the critical point $h_c=2$  for $D = 2$. 
 (d) Phase diagram of the transverse-field Ising model with DM interaction in the $h$--$D$ plane. (e) Defect density $n_d$ as a function of the ramp time scale $\tau$ in the TFIMDM. The system is quenched from $h_i = 8$ to $h_f = 1.5,\,0.5,\,-0.5,$ and $-1.5$, with system size $N = 1024$.}
 
\label{fig2}
\end{figure*}
%%%%%%%%%%%%%%%%%%%%%%%%%%%%%%%%%%%%%%%%%%%%%%%%%%%%%%%%%%%%%%%

We assume that the system initially prepared in its ground state $|\Psi_0\rangle$ at $\theta_i=0$, and here, we focus on a linear ramp of $\theta(t)=\pi t/\tau$, from an initial value $\theta_i=0$ at $t_i=0$ to the final value
$\theta_f=\pi$ at $t_f=\tau$ which crosses QCP at $\theta_c=\pi/2$. 
It is worthwhile to mention that, quench of $\theta(t)$ 
interpreting as a quench of an Aharonov-Bohm flux piercing the ring~\cite{Nakagawa2016,Jafari2021}.
When $\theta$
is ramped up to $\pi$ the state of the system gets excited from its instantaneous ground state, and its final
state at $t_f$ has a finite number of defects which is given by 
%
%%%%%%%%%%%%%%%%%%%%%%%%%%%%%%%%%%%  Eq.~FBH %%%%%%%%%%%%%%%%%%%%%%%%%%%%%%
\bea
\label{kink}
n_{d}=\frac{1}{N}\sum_{k}\sum_{\ell=\alpha,\beta}\langle\Psi_{0,k}(t_f)|\gamma_{k}^{{\ell}\dag}(t_f)\gamma_{k}^{{\ell}}(t_f)|\Psi_{0,k}(t_f)\rangle,
\quad
\eea
%%%%%%%%%%%%%%%%%%%%%%%%%%%%%%%%%%%%%%%%%%%%%%%%%%%%%%%%%%%%%%%%%%%%%%%
%
where $|\Psi_{0,k}(t_f)\rangle$ is the time-evolved state of $|\Psi_{0,k}(0)\rangle$ at $t_f$, and
$\gamma_{k}^{{\ell}}(t)=a^{{\ell}}_{k,q}(t)c^{q}_{k}+b^{{\ell}}_{k,q}(t)c^{q\dag}_{-k}+a^{{\ell}}_{k,p}(t)c^{p}_{k}+b^{{\ell}}_{k,p}(t)c^{p\dag}_{-k}$.
The coefficients $a^{{\ell}}_{k,j}(t)$ and $b^{{\ell}}_{k,j}(t)$ ($j=q,p$) obtained from solving the time dependent Schrödinger equation
%
%%%%%%%%%%%%%%%%%%%%%%%%%%%%%%%%%%%  Eq.~TDSE %%%%%%%%%%%%%%%%%%%%%%%%%%%%%%
\bea
\label{eqTDSE}
\left(
\begin{array}{c}
\dot{a}^{{\ell}}_{k,q}(t) \\
\dot{b}^{{\ell}}_{k,q}(t) \\
\dot{a}^{{\ell}}_{k,p}(t) \\
\dot{b}^{{\ell}}_{k,p}(t) \\
\end{array}
\right)
=
{\cal H}_k(t)
\left(
\begin{array}{c}
a^{{\ell}}_{k,q}(t) \\
b^{{\ell}}_{k,q}(t) \\
a^{{\ell}}_{k,p}(t) \\
b^{{\ell}}_{k,p}(t) \\
\end{array}
\right).
\eea
%%%%%%%%%%%%%%%%%%%%%%%%%%%%%%%%%%%%%%%%%%%%%%%%%%%%%%%%%%%%%%%%%%%%%%%
%
It should be noted that the density of defects is also given as $n_{d}=\frac{1}{N}\sum_{k}\sum_{\ell=\alpha,\beta}p^{\ell}_k$,
where $p^{\ell}_k$ is the transition probability of quasi-particle $\ell$ to the upper level which can also be calculated numerically
using the Von-Neumann equation~\cite{Jafari2025SM}.
The results of numerical calculations of defect density versus $\tau$ are remarkable as depicted in Fig.~\ref{fig1}(e)
for $J_e=1$ and different values of $J_o$. 
The surprising result arises at the IP ($J_o/J_e=1$), where the defect density deviates pronouncedly from the KZ scaling.
According to KZ scaling it is anticipated that, a large ramp time scale leads to lower non-adiabaticity, while at the IP
the density of defects enhances by increasing the ramp time scale which is indicative of AKZ behaviour. 
We have also found an intriguing and counterintuitive phenomenon away from the IP ($J_o/J_e\neq 1$). As seen, although the quench of magnetic flux crosses 
the critical point ($\theta_c=\pi/2$), the density of defects exhibits suppression faster than that predicted by the KZM.
%\\

To understand the origin of the different behaviors of $n_d$ at the IP and away from the IP, recall from  
Eq.~(\ref{kink}) that the defect density arises from the excitations of two distinct quasi-particles ($\alpha,\beta$).
The numerical simulation manifest that, the excitation of quasi-particle in the second filled band, $\varepsilon^{\beta}_k$, 
which controls the phase transitions (collapsing to zero for any values of $J_o/J_e$ at the critical point $\theta_c=\pi/2$ for all modes [Fig.~\ref{fig1}(a) and Fig.~\ref{fig1}(c)], vanishes as $\tau$ increases, i.e., $p^{\beta}_k \approx 0$.
Therefore, the dynamics of the system and consequently, the density of defect are managed by excitations of quasi-particles in the lowest energy band $\varepsilon^{\alpha}_k$.

As the $\varepsilon^{\alpha}_k$ band is gapless at $\theta_c=\pi/2$  ($k=\pm\pi$) only at the IP [Fig.~\ref{fig1}(a) and \ref{fig1}(c)], 
the system undergoes non-adiabatic evolution which allows quasi-particles to be easily excited to the upper state ($p^{\alpha}_k\neq 0$) resulting in a considerable defect density at the IP.
The AKZ behaviour of the defect density at the IP, might be attributed to supplementary excitation facilitated by the quasi-particle at the dispersionless band $\varepsilon^{\beta}_k$, at QCP.
However, away from the IP, while $\varepsilon^{\beta}_k$ band remains dispersionless at the QCP, the $\varepsilon^{\alpha}_k$ band (which controls the system's dynamics) is gapped for all modes $k$ [Figs. \ref{fig1}(b) and \ref{fig1}(d)], even at $\theta_c=\pi/2$, thus effectively hinders quasi-particle excitations from that band. 
Accordingly, while the quench crosses the critical point away from the IP, the density of excitations experiences a more significant suppression than anticipated by the KZM.

Following the demonstration that quantum criticality is not a sufficient condition for the emergence of Kibble-Zurek  scaling, we now turn to a complementary perspective. In particular, we show that KZ scaling of the defect density may persist even when the quench protocol crosses a non-critical point. Representative examples of this behavior include quenches across non-critical points in the transverse field Ising model with Dzyaloshinskii-Moriya interaction~\cite{Antal1997,Jafari2008,Wang2018} and in the generalized XY model~\cite{Titvinidze2003,Zvyagin2006,Divakaran2013}.
\\

{\em Transverse field Ising model with DM interaction (TFIMDM)} $-$
The Hamiltonian of TFIMDM
%transverse field Ising model with DM interaction
is given as~\cite{Antal1997,Jafari2008,Wang2018}
%
%%%%%%%%%%%%%%%%%%%%%%%%%%%%%%%%%%%  TFIDM %%%%%%%%%%%%%%%%%%%%%%%%%%%%%%
\begin{equation}
\label{TFIDM}
{\cal H}^{\rm DM}
\!=\!
 -\frac{J}{2}\sum_{n=1}^N\Big[\sigma_n^x \sigma^x_{n+1} + \frac{D}{2}  (\sigma_n^x \sigma^y_{n+1} -  \sigma_n^y \sigma^x_{n+1}) - h \sigma_n^z \Big],
\end{equation}
%%%%%%%%%%%%%%%%%%%%%%%%%%%%%%%%%%%%%%%%%%%%%%%%%%%%%%%%%%%%%%%%%%%%%%%%%%%
%
with $s_n^{\delta}=\sigma^{\delta}_n/2$. 
Here $\sigma^{\delta}_n$ ($\delta=x,y,z$) are Pauli matrices acting at sites $n$ of a one-dimensional lattice.
We consider the periodic boundary condition and without loss of generality we set $J=1$ as the energy scale.
The Hamiltonian ${\cal H}^{\rm DM}$ %in Eq.~(\ref{TFIDM}) 
is still exactly solvable in the presence of DM interaction through the Jordan-Winger and Fourier transformations~\cite{Antal1997,Wang2018,Derzhko2006}, which are expressed as a sum over decoupled mode Hamiltonians ${\cal H}^{\rm DM}_{k}$ by introducing 
the Nambu spinors ${\mathbb C}_k^{\dagger} = (c_k^{\dagger} \ c_{-k})$, 
%
%%%%%%%%%%%%%%%%%%%%%%%%%%%%%%%%%%%%%%%%%  Eq.~Nambu  %%%%%%%%%%%%%%%%%%%%%%%%%%%%%%%%%%%%%%%%%%%
\bea
\label{eq:Nambu}
{\cal H}^{\rm DM} = \sum_{k>0} {\mathbb C}^{\dagger}_k {\mathcal H}_{k}^{\rm DM} {\mathbb C}_k, 
\eea
%%%%%%%%%%%%%%%%%%%%%%%%%%%%%%%%%%%%%%%%%%%%%%%%%%%%%%%%%%%%%%%%%%%%%%%%%%%%%%%%%%%%
%
with $ k= \frac{(2m-1)\pi}{N}$, $m=1,\cdots,N/2$, and 
%
%%%%%%%%%%%%%%%%%%%%%%%%%%%%%%%%%%%%%%%%%  Eq.~Nambu  %%%%%%%%%%%%%%%%%%%%%%%%%%%%%%%%%%%%%%%%%%%
\begin{equation}
\label{eq:matrix}
{\cal H}_{k}^{\rm DM}=\left(
                       \begin{array}{cc}
                         h-\cos(k)-D\sin(k) & \sin(k) \\
                         \sin(k) & -h+\cos(k)-D\sin(k) \\
                       \end{array}
                     \right),
\end{equation}
%%%%%%%%%%%%%%%%%%%%%%%%%%%%%%%%%%%%%%%%%%%%%%%%%%%%%%%%%%%%%%%%%%%%%%%%%%%%%%%%%%%%
%where $\mathbb{1}$ denotes the $2\times 2$ identity matrix. 
in which the eigenvalues of Hamiltonian ${\mathcal H}_{k}^{\rm DM}$ are $\varepsilon^{\pm}_k=-D\sin(k)\pm\sqrt{(h-\cos(k))^2+\sin^2(k)}$.
The spectrum of the model is shown in Fig.~\ref{fig2}(a)-(c) for $h=1$ and $h=2$ with different values of the DM interaction. 
It is evident that the $k \rightarrow -k$ symmetry of the spectrum is broken in the presence of the DM interaction. 
In one-dimensional electron or spin systems, quantum phase transitions are often associated with changes in the topology of the Fermi surface, 
which may occur with or without gap closing, and are typically characterized by the doubling of the number of Fermi points~\cite{Antal1997,Fabrizio1996,Titvinidze2003,Arita1998,Daul2000,Aebischer2001}.

The phase diagram of the TFIMDM has been depicted in Fig.~\ref{fig2}(d).
It has been verified that, for small DM interaction ($D<1$), the ground state remains that of the transverse Ising model ($D=0$) 
with no energy current flows~\cite{Antal1997}.
In other words, when $D<1$ the critical points of system are still located at the gap closing points $h_c=\pm1$, correspond to the QPTs
from a ferromagnetically ordered phase for $|h|<1$ to the paramagnetic phase ($|h|>1$) with the associated exponents
being the same as the transverse filed Ising model~\cite{Antal1997}. However, the topology of the Fermi surface and consequently 
the ground-state properties do change for $D>1$, with the nonzero energy current~\cite{Antal1997}. As a result, when $D>1$ a new critical line, $h_c=D$ appears  
where the topology of the Fermi surface changes by lowering the number of Fermi point from four for $h<h_c=D$ to none for $h>h_c=D$  \cite{Antal1997,Derzhko2006,Wang2018,Haldar2020,Roy2019}.
%, as shown in   Fig. \ref{fig2}(b). 
%

In our analysis, we consider the linear ramp of transverse field characterized by $h(t)=t/\tau$, with $t \in [h_i\tau, h_f\tau]$. The excitation probability $p_k$ and consequently the density of defects $n_d=\frac{1}{N}\sum_k p_k$ can be calculated analytically by solving the time dependent Schrödinger equation as well as numerically from the Von-Neumann equation~\cite{Jafari2025SM}. As expected, our numerical simulation exhibits that when $D<1$ for a quench that crosses the Ising type QCP $h_c=1$, the density of defects manifests KZ scaling $n_d\propto\tau^{-\beta}$ with $\beta=d\nu/(1+z\nu)=1/2$, where $\nu=1$,  $z=1$ and $d=1$~\cite{Dziarmaga2005,Jafari2025SM}.

For a DM interaction strength $D=2$, the dependence of the excitation density on the ramp time scale $\tau$ is illustrated in Fig.~\ref{fig2}(e) for quenches starting from $h_i=8$ and ending at different values of the final field $h_f$. As shown, when the quench crosses the critical point at $h_c=2$, specifically for $h_f=1.5$, the resulting defect density exhibits a markedly faster suppression with increasing $\tau$ than that predicted by the Kibble-Zurek mechanism.
Furthermore, the KZ scaling $n_d \propto \tau^{-1/2}$ persists even for quenches across the {\em non-critical} point at $h=1$, with final field values $h_f=0.5$, $-0.5$, and $-1.5$.
 %(Fig.~\ref{fig2}(e)). 
Based on our findings for the GCM, we address this conundrum by examining the quasiparticle spectrum of the model, as shown in Fig.~\ref{fig2}(a)-(c).

Analogous to the GCM away from the isotropic point, the defect density shows an accelerated suppression relative to the Kibble-Zurek scaling when the quench crosses the critical point at $h_c=2$ for $D=2$. This behavior originates from the fact that the quasiparticles governing the dynamics remain fully gapped even at the quantum critical point, as shown in Fig.~\ref{fig2}(c), thereby rendering the passage across $h_c=2$ effectively adiabatic.
In contrast, when the quench crosses the {\em non-critical} point at $h=1$ for $D=2$, the defect density follows a power-law scaling consistent with the Kibble-Zurek prediction. This scaling can be attributed to the presence of massless quasiparticles, as illustrated in Fig.~\ref{fig2}(b). Notably, this mechanism parallels that of the transverse field Ising model with weak DM interaction ($D<1$) at an Ising-like transition point~\cite{Dziarmaga2005,Jafari2025SM}, but here it arises for a quench through a non-critical point, as shown in Fig.~\ref{fig2}(b).
%\\

This picture is further corroborated by our analysis of the GXY model, discussed in~\cite{Jafari2025SM}. When the system is quenched through a critical point, the defect density decreases with increasing ramp time scale but departs from the standard Kibble-Zurek scaling, exhibiting a significantly faster suppression. In contrast, quenches through a {\em non-critical} point recover the expected Kibble-Zurek scaling behavior. This dichotomy can be understood in terms of the quasiparticle spectrum: the anomalously fast decay arises from gapped quasiparticles persisting at the critical point, whereas the restoration of Kibble-Zurek scaling for non-critical quenches is due to the presence of massless quasiparticles~\cite{Jafari2025SM}.
\\

{\em Summary and conclusion} $-$ 
In conclusion, we have shown that the presence of a quantum phase transition is neither a necessary nor a sufficient condition for Kibble–Zurek scaling and non-adiabatic dynamics. 
What ultimately determines defect generation is the gap structure of the quasiparticle modes that govern the excitation probabilities during the ramp. 
This dynamical property may or may not coincide with the equilibrium critical point.

In the generalized compass model, a suppression faster than the conventional Kibble–Zurek scaling is observed away from the isotropic point, 
where defect generation is controlled by quasiparticle states that remain gapped at the anisotropic quantum phase transition. 
A similar accelerated suppression occurs in the transverse-field Ising model with Dzyaloshinskii–Moriya interaction, 
where quenches across the critical line involve dynamically gapped quasiparticles. 
In contrast, quenches through a {\em non-critical} point recover Kibble–Zurek scaling, 
reflecting the presence of gapless quasiparticles that actively participate in the dynamics. 
Our analysis of the generalized XY model further supports this unified picture, 
again revealing rapid suppression at critical points and conventional scaling at non-critical points.

Taken together, these results demonstrate that the scaling of defect production is dictated by the quasiparticles responsible for the system's dynamics, which may remain either gapless or gapped at a quantum critical point. 
Our findings provide a refined understanding of the relation between defect scaling and quantum criticality, and offer new insight into non-adiabatic dynamics and strategies for achieving controlled adiabatic evolution in driven quantum systems.
\\

{\it Acknowledgements} $-$
A.A. is supported by the Beijing Natural Science Foundation (Grant No. IS25015).

%\bibliography{AKZ_References}

%

\vskip 15 cm
\begin{widetext}

\setcounter{figure}{0}
\setcounter{equation}{0}
\setcounter{section}{0}

\newpage

\section{Supplementary material}
\renewcommand\thefigure{S\arabic{figure}}
\renewcommand\theequation{S\arabic{equation}}

\vskip 0.5 cm

In this Supplemental material we elaborate on some technical aspects of the analysis presented in the accompanying Letter \cite{Jafari2025b}, and also provide some background material.

\subsection{A. Generalized Compass Model (GCM)}
The Hamiltonian of the 1D spin-$1/2$ generalized quantum compass model (GCM) is given by \cite{You2014SM}
%
%%%%%%%%%%%%%%%%%%%%%%%%%%%%%%%%%%%%%% Eq. S1 %%%%%%%%%%%%%%%%%%%%%%%%%%%%%%%%%%%%
\bea
\label{eqS1}
{\cal H}=
-\sum_{n=1}^{N}
\Big[J_o\tilde{\sigma}_{2n-1}(\theta)\tilde{\sigma}_{2n}(\theta)
+ J_e\tilde{\sigma}_{2n}(-\theta)\tilde{\sigma}_{2n+1}(-\theta) \Big],
\eea
%%%%%%%%%%%%%%%%%%%%%%%%%%%%%%%%%%%%%%%%%%%%%%%%%%%%%%%%%%%%%%%%%%%%%%%%%%%%%%%%%%
%
with $J_{e/o}$ exchange amplitudes on even/odd lattice bonds, and where the pseudo-spin operators $\tilde{\sigma}_n(\pm \theta)$ are
formed by linear combinations of the Pauli matrices $\sigma_{n}^{x}$ and $\sigma_{n}^{y}$: $\tilde{\sigma}_n(\pm \theta) =\cos(\theta) \sigma_{n}^{x} \mp \sin(\theta) \sigma_{n}^{y}$. This Hamiltonian can be diagonalized exactly by mapping it onto a free fermion model,
%
%%%%%%%%%%%%%%%%%%%%%%%%%%%%%%%%%%%%%% Eq. S2 %%%%%%%%%%%%%%%%%%%%%%%%%%%%%%%%%%%%
\bea
\label{eqS2}
{\cal H}=-\sum_{n=1}
\Big[(J_{o} c^{\dagger}_{2n}c_{2n-1}+J_{e} c^{\dagger}_{2n+1}c_{2n}+\mbox{H.c.})
+(J_{o} e^{-i\theta} c^{\dagger}_{2n}c^{\dagger}_{2n-1}+J_{e} e^{i\theta}c^{\dagger}_{2n+1}c^{\dagger}_{2n}+\mbox{H.c.})\Big],
\eea
%%%%%%%%%%%%%%%%%%%%%%%%%%%%%%%%%%%%%%%%%%%%%%%%%%%%%%%%%%%%%%%%%%%%%%%%%%%%%%%%%%
%
using the Jordan-Wigner transformation
%
%%%%%%%%%%%%%%%%%%%%%%%%%%%%%%%%%%%%%% Eq. S3 %%%%%%%%%%%%%%%%%%%%%%%%%%%%%%%%%%%%
\bea
\label{JWT}
s^{+}_{n}=s^{x}_{n} + {\it i}s^{y}_{n}= \prod_{m=1}^{n-1}\left(1-2c_{m}^{\dagger}c_{m}\right)c_{n}^{\dagger},~~
s^{-}_{n}=s^{x}_{n} - {\it i}s^{y}_{n}= \prod_{m=1}^{n-1}c_{n}\left(1-2c_{m}^{\dagger}c_{m}\right),~~
s^{z}_{n}= c_{n}^{\dagger}c_{n}-1/2.
\eea
%%%%%%%%%%%%%%%%%%%%%%%%%%%%%%%%%%%%%%%%%%%%%%%%%%%%%%%%%%%%%%%%%%%%%%%%%%%%%%%%%%
%
By partitioning the chain into bi-atomic
elementary cells and defining two independent fermions at each cell $n$  \cite{Perk1975SM,Derzhko2009SM}, $c_{n}^{q}\equiv c_{2n-1}$
and $c_{n}^{p}\equiv c_{2n}$, one can rewrite the Hamiltonian in Eq. (\ref{eqS2}) as
%
%%%%%%%%%%%%%%%%%%%%%%%%%%%%%%%%%%%%%% Eq. S4 %%%%%%%%%%%%%%%%%%%%%%%%%%%%%%%%%%%%
\bea
\label{eqS4}
{\cal H}=-\sum_{n=1}^{N/2}
\Big[
J_{o}c^{q\dagger}_{n}c^{p}_{n}+J_{e} c^{q\dagger}_{n+1}c^{p}_{n}+\mbox{H.c.})
+(J_{o} e^{-i\theta} c^{q\dagger}_{n}c^{p\dagger}_{n}+J_{e} e^{i\theta}c^{q\dagger}_{n+1}c^{p\dagger}_{n}+\mbox{H.c.})\Big].
\eea
%%%%%%%%%%%%%%%%%%%%%%%%%%%%%%%%%%%%%%%%%%%%%%%%%%%%%%%%%%%%%%%%%%%%%%%%%%%%%%%%%%
%

By Fourier transforming the GCM Hamiltonian ${\cal H}$ in Eq. (\ref{eqS4}), and grouping together terms with $k$ and $-k$, ${\cal H}$ 
is transformed into a sum of commuting Hamiltonians ${\cal H}_k$, i.e., ${\cal H}= \sum_{k\ge0}C^{\dagger} {\cal H}_{k} C$ each describing a different $k$ mode,
%
%%%%%%%%%%%%%%%%%%%%%%%%%%%%%%%%%%%%%% Eq. S5 %%%%%%%%%%%%%%%%%%%%%%%%%%%%%%%%%%%%
\bea
\label{eqS5}
{\cal H}_k= J_{k}c_{k}^{q\dag}c_{-k}^{p\dag} +L_{k}c_{k}^{q\dag}c_{k}^{p}
+ J_{-k}c_{-k}^{q\dag}c_{k}^{p\dag} + L_{-k} c_{-k}^{q\dag}c_{-k}^{p} + \mbox{H. c.}
\eea
%%%%%%%%%%%%%%%%%%%%%%%%%%%%%%%%%%%%%%%%%%%%%%%%%%%%%%%%%%%%%%%%%%%%%%%%%%%%%%%%%%
%
Here $L_{k}\!=\!(J_o+J_e e^{ik})$ and $J_{k}\!=\!(J_o e^{i\theta}-J_e e^{i(k-\theta)})$.
We can thus obtain the spectrum of the GCC model by diagonalizing each Hamiltonian mode ${\cal H}_k$ in (\ref{eqS5}) independently. This can be done in two ways: Using a generalized Bogoliubov transformation which maps ${\cal H}_k$ onto the BdG quasiparticle Hamiltonian ${\cal H}(k)$ in Eq. (4) in \cite{Jafari2025b}  (with the quasiparticle operators $\gamma_k^{\ell}$ and $\gamma_k^{\ell \dagger}$ ($\ell=\alpha,\beta$), expressed in terms of the fermion operators in (\ref{eqS5})), or using a basis in which the eigenstates of ${\cal H}_k$ are obtained as linear combinations of even-parity fermion states \cite{Sun2009SM}. Here we outline the connection between the two approaches.

As ${\cal H}_k$ in (\ref{eqS5}) conserves the number parity (even or odd number of fermions), it is sufficient to consider the even-parity subspace of the Hilbert space. This subspace is spanned by the eight basis vectors
%
%%%%%%%%%%%%%%%%%%%%%%%%%%%%%%%%%%%%%% Eq. S6 %%%%%%%%%%%%%%%%%%%%%%%%%%%%%%%%%%%%
\bea
\label{eqS6}
\bl
|\varphi_{1,k}\rangle&=|0\rangle &
|\varphi_{2,k}\rangle&=c_{k}^{q\dag}c_{-k}^{q\dag}|0\rangle, &
|\varphi_{3,k}\rangle&=c_{k}^{q\dag}c_{-k}^{p\dag}|0\rangle, &
|\varphi_{4,k}\rangle&=c_{-k}^{q\dag}c_{k}^{p\dag}|0\rangle,  \\
|\varphi_{5,k}\rangle&=c_{k}^{q\dag}c_{-k}^{p\dag}|0\rangle, &
|\varphi_{6,k}\rangle&=c_{k}^{q\dag}c_{-k}^{q\dag}c_{k}^{p\dag}c_{-k}^{p\dag}|0\rangle, &
|\varphi_{7,k}\rangle&=c_{k}^{q\dag}c_{k}^{p\dag}|0\rangle, &
|\varphi_{8,k}\rangle&=c_{-k}^{q\dag}c_{-k}^{p\dag}|0\rangle,
\el
\eea
%%%%%%%%%%%%%%%%%%%%%%%%%%%%%%%%%%%%%%%%%%%%%%%%%%%%%%%%%%%%%%%%%%%%%%%%%%%%%%%%%%%
%
where, $|0\rangle$ denotes the vacuum state of the Fermi operators $c_{k}^{p,q}$.
Due to the total momentum conservation, the above sets of basis block diagonalize ${\cal H}_k$ into two blocks of dimensions
$6$ and $2$ such that ${\cal H}_k=\bigoplus_{d=2,6}{\cal H}_k^{(d)}$ with
%
%%%%%%%%%%%%%%%%%%%%%%%%%%%%%%%%%%%%%% Eq. S7 %%%%%%%%%%%%%%%%%%%%%%%%%%%%%%%%%%%%
\bea
\bl
\label{eqS7}
{\cal H}_k^{(6)}=
\left(
  \begin{array}{cccccc}
    0 & 0 & J_k^{\ast} & J_{-k}^{\ast} & 0 & 0 \\
    0 & 0 & L_k^{\ast} & -L_k & 0 & 0 \\
    J_k & L_k & 0 & 0 & L_k & J_{-k}^{\ast} \\
    J_{-k} & -L_k^{\ast} & 0 & 0 & -L_k^{\ast} & J_k^{\ast} \\
    0 & 0 & L_k^{\ast} & -L_k & 0 & 0 \\
    0 & 0 & J_{-k} & J_k & 0 & 0 \\
  \end{array}
\right);
\el
\quad
\quad
{\rm and}
\quad
\quad
\label{eqS8}
{\cal H}_k^{(2)}=
\left(
  \begin{array}{cc}
    0 & 0 \\
    0 & 0 \\
  \end{array}
\right).
\eea
%%%%%%%%%%%%%%%%%%%%%%%%%%%%%%%%%%%%%%%%%%%%%%%%%%%%%%%%%%%%%%%%%%%%%%%%%%%%%%%%%%%
%
The eigenstates $|\psi_{m,k}\rangle$ of ${\cal H}_k^{(6)}$ in this basis can be written as
%
%%%%%%%%%%%%%%%%%%%%%%%%%%%%%%%%%%%%%% Eq. S9 %%%%%%%%%%%%%%%%%%%%%%%%%%%%%%%%%%%%
\bea
\label{eqS9}
|\psi_{m,k}\rangle
=\sum_{j=1}^{6}v_{m,k}^{j}|\varphi_{j,k}\rangle,
\eea
%%%%%%%%%%%%%%%%%%%%%%%%%%%%%%%%%%%%%%%%%%%%%%%%%%%%%%%%%%%%%%%%%%%%%%%%%%%%%%%%%%%
%
where $|\psi_{m,k}\rangle$ is an unnormalized eigenstate of ${\cal H}_k^{(6)}$
with corresponding eigenvalue $\epsilon_{m,k}\, (m=1,\cdots,6)$, and where
$v_{m,k}^{j}\, (j=1,\cdots,6)$ are functions of $J_e,J_o$) and $\theta$, and the momentum $k$.
Two eigenstates are degenerate with eigenvalues zero ($\epsilon_{3,k}=\epsilon_{4,k}=0$), with the ground state and the first excited state having negative energies
($\epsilon_{1,k}=-\epsilon_{6,k}=(\varepsilon_{k}^{\alpha}+\varepsilon_{k}^{\beta}), \epsilon_{2,k}=-\epsilon_{5,k}=(\varepsilon_{k}^{\alpha}-\varepsilon_{k}^{\beta})$ respectively). Here $\varepsilon_{k}^\alpha$ and $\varepsilon_{k}^\beta$ are the quasiparticle energies defined after Eq. (4) in \cite{Jafari2025b}.
Two eigenvalues of Hamiltonian ${\cal H}_k^{(2)}$ are $\epsilon_{7,k}=\epsilon_{8,k}=0$, with corresponding eigenvectors $|\varphi_{7,k}\rangle$ and $|\varphi_{8,k}\rangle$. 
Each eigenstate of ${\cal H}_k$ can be linked to a state in the BdG formalism via their common eigenvalues.
For instance, the ground state $|\psi_{1,k}\rangle$ of ${\cal H}_k$ is identified with the BdG mode with the corresponding negative energy quasiparticle states filled,
i.e. $|\Psi_{0,k}\rangle = \gamma^{2\dagger}_{k}\gamma^{1\dagger}_{k}|V_k\rangle=|\psi_{1,k}\rangle$, where $|V_k\rangle$ is the $k^{\text th}$ single-fermion mode of the Bogoliubov vacuum.

Due to the block-diagonal form of ${\cal H}_k$, for the time dependent $\theta(t)=\pi t/\tau$ ($t \in[0,\tau]$), the dynamics of system govern by two decoupled Von-Neumann equations 
$i {\dot{\rho}_k}^{(d)}(t)=[{\cal H}_k^{(d)}(t),\rho_k^{(d)}(t)]$. Therefore, the density of defects can be calculated numerically as 
\bea
n_d=1-\frac{1}{N}\sum_k\langle\psi_{1,k}(t_f)|\rho^{(6)}_k(t_f)|\psi_{1,k}(t_f)\rangle
\eea
where $|\psi_{1,k}(t_f)\rangle=|\psi_{1,k}(\theta_f)\rangle$ is the ground state of the system at $t_f=\tau$.

Let us point out that while the BdG formalism is very convenient for obtaining energy eigenvalues, the fermionic even-parity basis is preferable
for numerically computing matrix elements of the time-evolved states, and consequently density of defects. 
The unnormalized ground state of the model at the IP $J_e=J_o=J$ is given as
%
%%%%%%%%%%%%%%%%%%%%%%%%%%%%%%%%%%%%%% Eq. S10 %%%%%%%%%%%%%%%%%%%%%%%%%%%%%%%%%%%%
\bea
\label{eqS10}
|\psi_{1,k}\rangle
=\sum_{j=1}^{6}v_{0,k}^{j}|\varphi_{j,k}\rangle,
\eea
%%%%%%%%%%%%%%%%%%%%%%%%%%%%%%%%%%%%%%%%%%%%%%%%%%%%%%%%%%%%%%%%%%%%%%%%%%%%%%%%%%%
%
with 
%
%%%%%%%%%%%%%%%%%%%%%%%%%%%%%%%%%%%%%% Eq. S11 %%%%%%%%%%%%%%%%%%%%%%%%%%%%%%%%%%%%
\bea
\label{eqS11}
\bl
v_{1,k}^{1}&=1,
\\
v_{1,k}^{2}&=-i \sec (\theta ) \cot \left(\frac{k}{2}\right),
\\
v_{1,k}^{3}&=\frac{1}{2} i \sec (\theta ) \left(\cot \left(\frac{k}{2}\right)+i\right) \sqrt{\cos (2 \theta )+2 \sin ^2(\theta ) \cos (k)+3},
\\
v_{1,k}^{4}&=-\frac{1}{2} i \sec (\theta ) \left(\cot \left(\frac{k}{2}\right)-i\right) \sqrt{\cos (2 \theta )+2 \sin ^2(\theta ) \cos (k)+3},
\\
v_{1,k}^{5}&=-i \sec (\theta ) \cot \left(\frac{k}{2}\right),
\\ 
v_{1,k}^{6}&=1,
\el
\eea
%%%%%%%%%%%%%%%%%%%%%%%%%%%%%%%%%%%%%%%%%%%%%%%%%%%%%%%%%%%%%%%%%%%%%%%%%%%%%%%%%%%
% 

and normalization constants $\Big(2 \left(\sec ^2(\theta )+\tan ^2(\theta ) \cos (k)+1\right)\Big)^{-1/2}$.\\

The unnormalized ground state of the model at the anisotropic point $J_e \neq J_o$ is given by Eq. (\ref{eqS10})
with 
%
%%%%%%%%%%%%%%%%%%%%%%%%%%%%%%%%%%%%%% Eq. S12 %%%%%%%%%%%%%%%%%%%%%%%%%%%%%%%%%%%%
\bea
\label{eqS12}
\bl
v_{1,k}^{1}&=\frac{J_o+e^{2 i \theta } J_e}{J_e+e^{2 i \theta } J_o},\\
%\no
v_{1,k}^{2}&=\frac{\cot \left(\frac{k}{2}\right) \left(J_e+J_o\right) (\sin (\theta )-i \cos (\theta ))}{J_e+e^{2 i \theta } J_o},\\
%\no
v_{1,k}^{3}&=-\frac{2 e^{i (\theta +k)} \sqrt{J_e J_o \left(2 \cos (2 \theta ) \sin ^2\left(\frac{k}{2}\right)+\cos (k)\right)+J_e J_o+J_e^2+J_o^2}}{\left(-1+e^{i k}\right) \left(J_e+e^{2 i \theta } J_o\right)},\\
%\no
v_{1,k}^{4}&=\frac{2 e^{i \theta } \sqrt{J_e J_o \left(2 \cos (2 \theta ) \sin ^2\left(\frac{k}{2}\right)+\cos (k)\right)+J_e J_o+J_e^2+J_o^2}}{\left(-1+e^{i k}\right) \left(J_e+e^{2 i \theta } J_o\right)},\\
%\no
v_{1,k}^{5}&=\frac{\cot \left(\frac{k}{2}\right) \left(J_e+J_o\right) (\sin (\theta )-i \cos (\theta ))}{J_e+e^{2 i \theta } J_o},\\
%\no 
v_{1,k}^{6}&=1,
\el
\eea
%%%%%%%%%%%%%%%%%%%%%%%%%%%%%%%%%%%%%%%%%%%%%%%%%%%%%%%%%%%%%%%%%%%%%%%%%%%%%%%%%%%
% 
and normalization constants 
$$
\Big(-\frac{16 e^{i (2 \theta +k)} \left(J_e J_o \left(2 \cos (2 \theta ) \sin ^2\left(\frac{k}{2}\right)+\cos (k)\right)+J_e J_o+J_e^2+J_o^2\right)}{\left(-1+e^{i k}\right)^2 \left(J_o+e^{2 i \theta } J_e\right) \left(J_e+e^{2 i \theta } J_o\right)}\Big)^{-1/2}
\!
=
 \!
\frac{\bigl|\sin\!\tfrac{k}{2}\bigr|\;\sqrt{\,J_e^2 + J_o^2 + 2 J_e J_o \cos(2\theta)\,}}
{2\,\sqrt{\,J_e^2 + J_o^2 + 2 J_e J_o \bigl(\cos^2\theta + \sin^2\theta \cos k\bigr)\,}} .
$$

\vskip 1.0 cm

\subsection{B. Transverse field Ising model with DM interaction (TFIMDM)}
The Hamiltonian of transverse field Ising model with Dzyaloshinsky-Moriya (DM) interaction is given as \cite{Antal1997SM,Jafari2008SM,Wang2018SM}
%
%%%%%%%%%%%%%%%%%%%%%%%%%%%%%%%%%%%  Eq. S13 %%%%%%%%%%%%%%%%%%%%%%%%%%%%%%
\bea
\label{eqS13}
{\cal H}^{DM}(t)= -\frac{J}{2}\sum_{n=1}^N\Big[\sigma_n^x \sigma^x_{n+1} + \frac{D}{2}  (\sigma_n^x \sigma^y_{n+1} -  \sigma_n^y \sigma^x_{n+1}) - h(t) \sigma_n^z \Big],
\eea
%%%%%%%%%%%%%%%%%%%%%%%%%%%%%%%%%%%%%%%%%%%%%%%%%%%%%%%%%%%%%%%%%%%%%%%%%%%
%
We consider the periodic boundary condition and without loss of generality we set $J=1$ as the energy scale.
The Hamiltonian Eq. (\ref{eqS13}) can be diagonalized by mapping the spins to spinless fermions through the Jordan-Winger
transformation
%
%%%%%%%%%%%%%%%%%%%%%%%%%%%%%%%%%%%  Eq. S14 %%%%%%%%%%%%%%%%%%%%%%%%%%%%%%
\bea
\label{eqS14}
{\cal H}^{DM}(t)=-\frac{J}{2}\sum_{n=1}^N\Big[(1+iD)c^{\dagger}_n c_{n+1} - (1-iD)c_n c_{n+1}^{\dagger} + (c^{\dagger}_n c_{n+1}^{\dagger}-
c_n c_{n+1})-2h(t)c^{\dagger}_n c_{n}\Big].
\eea 
%%%%%%%%%%%%%%%%%%%%%%%%%%%%%%%%%%%%%%%%%%%%%%%%%%%%%%%%%%%%%%%%%%%%%%%%%%%
%
By Fourier transforming the GCM Hamiltonian ${\cal H}^{DM}$ in Eq. (\ref{eqS14}), and grouping together terms with $k$ and $-k$, ${\cal H}^{DM}(t)$ 
is transformed into a sum of commuting Hamiltonians ${\cal H}^{DM}_k(t)$, i.e., ${\cal H}^{DM}(t)= \sum_{k\ge0}{\mathbb C}_k^{\dagger} {\cal H}_{k}^{DM}(t) {\mathbb C}_k$ 
with ${\mathbb C}_k^{\dagger} = (c_k^{\dagger} \ c_{-k})$, and
%
%%%%%%%%%%%%%%%%%%%%%%%%%%%%%%%%%%%%%%%%%  Eq. S15  %%%%%%%%%%%%%%%%%%%%%%%%%%%%%%%%%%%%%%%%%%%
\bea
\label{eqS15}
{\cal H}_{k}^{DM}(t)=\left(
                       \begin{array}{cc}
                         h(t)-\cos(k) & \sin(k) \\
                         \sin(k) & -h(t)+\cos(k) \\
                       \end{array}
                     \right)-D\sin(k)\mathbb{1}.
\eea
%%%%%%%%%%%%%%%%%%%%%%%%%%%%%%%%%%%%%%%%%%%%%%%%%%%%%%%%%%%%%%%%%%%%%%%%%%%%%%%%%%%%
%
For linear time dependent magnetic field $h(t)=t/\tau$, the Hamiltonian in Eq. (\ref{eqS15}) for each mode has the form
${\cal H}_{k}^{DM}(t)=\tau_k/\tau \sigma^x+\Delta_k \sigma^z$, with $\tau_k=(h(t)-\cos(k))\tau$ and $\Delta_k=\sin(k)$, so transition rates can be calculated
by the Landau-Zener formula \cite{ZenerSM,LandauSM,Vitanov1999SM}.
The density of defects can also be calculated numerically as $n_d=1-\frac{1}{N}\sum_k\langle\psi_{g,k}(t_f)|\rho_k(t_f)|\psi_{g,k}(t_f)\rangle$
where $|\psi_{g,k}(t_f)\rangle$ is the ground state of the Hamiltonian at $t_f$, and the density matrix $\rho(t)$ is calculated numerically using the Von-Neumann equation $i {\dot{\rho}_k}(t)=[{\cal H}_k^{DM}(t),\rho_k(t)]$.\\

In Fig. \ref{figSM1}(a)-(b) the transition probability has been shown versus $k$ for a quench from $h_i=8$ to the different values of quench end field $h_f=1.5, 0.5, -0.5$ and $-1.5$ 
for $\tau=1$, and for $D=0.5$ and $D=2$. As predicted, for $D=1/2$ (Fig. \ref{figSM1}(a)), the quench does not cross the critical points $h_c=\pm 1$ ($k=0, \pi$) for $h_f=1.5$, the excitation probability is very small. However, for a quench filed end $h_f=0.5$, and $-0.5$ where the quench crosses the single critical point $h_c=1$, at $k=0$ 
the transition probability is maximum at the gap closing mode $p_{k=0}=1$. While away from the gap closing mode the system evolves adiabatically and $p_{k\rightarrow\pi}\rightarrow0$. In addition, for $h_f=1.5$, the quench crosses two critical points ($h_c=\pm1$) which result in maximum transition probability at
the gap closing modes $k=0, \pi$. 

For $D>1$, the critical point of the system is $h_c=D$ \cite{Antal1997SM,Derzhko2006SM,Wang2018SM,Haldar2020SM,Roy2019SM}. The transition probability has been plotted
in Fig. \ref{figSM1}(b), for $D=2$ ($h_c=\pm2$). As seen, the transition probability is very small for a quench across the critical point $h_c=2$ ($h_f=1.5$) while the transition probability is significant for a quench across the non-critical points $h=\pm1$ at modes $k=0, \pi$ where the quasi-particle band becomes gapless.\\

The density of defects in transverse field Ising model with Dzyaloshinsky-Moriya (DM) has been depicted in Fig. \ref{figSM1}(c) for a quench starts from $h_i=8$ to different values of quench field end $h_f$, for $D=1/2$ . As expected, our numerical simulation exhibits that for $D<1$ for a quench that crosses the Ising type QCP $h_c=1$, ($h_f=0.5, -0.5, -1.5$), the density of defects manifests KZ scaling $n_d\propto\tau^{-\beta}$ with $\beta=d\nu/(1+z\nu)=1/2$ where $\nu=1$,  $z=1$ and $d=1$ \cite{Dziarmaga2005sm}.
%
%%%%%%%%%%%%%%%%%%%%%%%%%%%%  Fig.SM1 %%%%%%%%%%%%%%%%%%%%%%%%%%%%%%%%
\begin{figure*}[t!]
\centerline{
\includegraphics[width=0.33\linewidth]{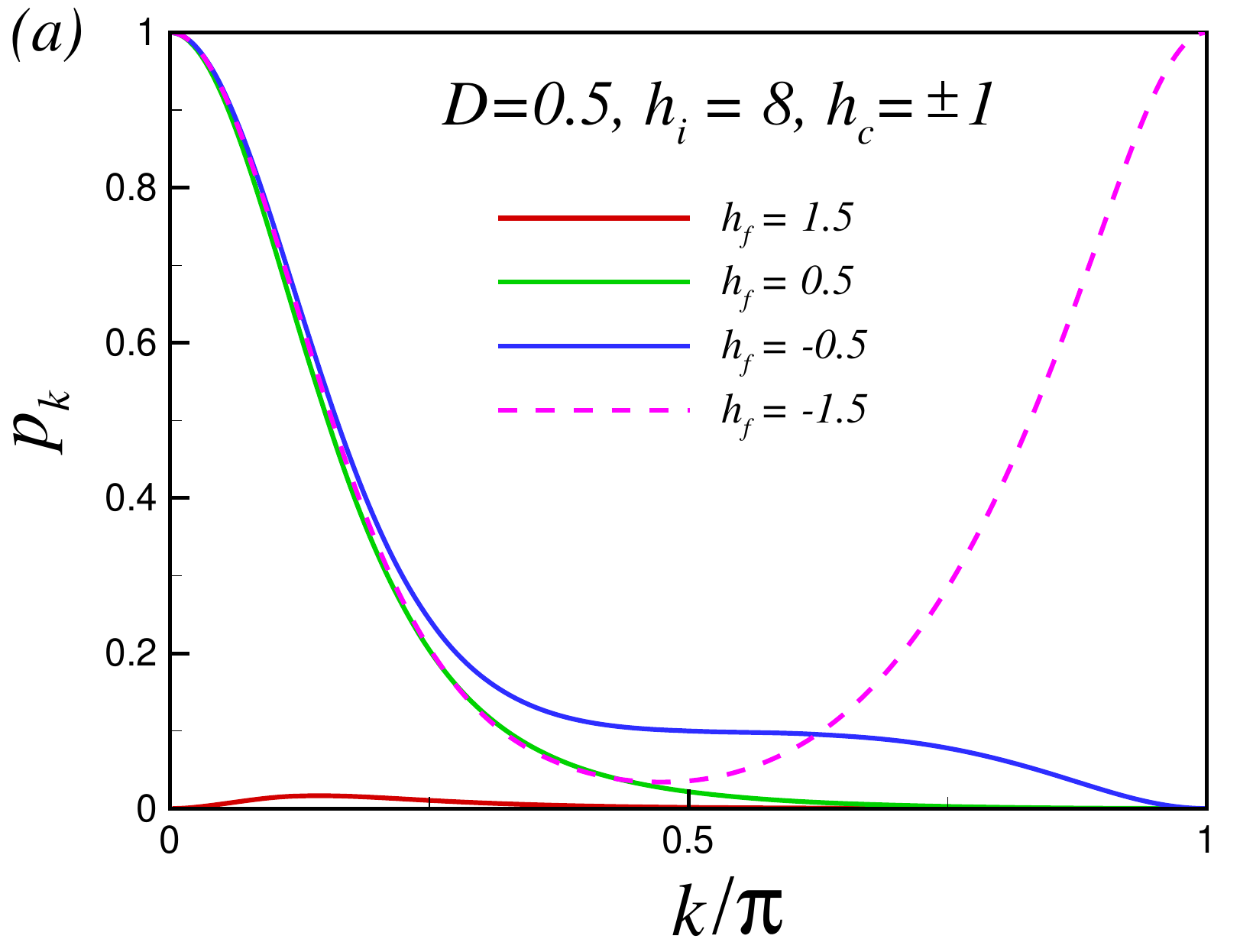}
\includegraphics[width=0.33\linewidth]{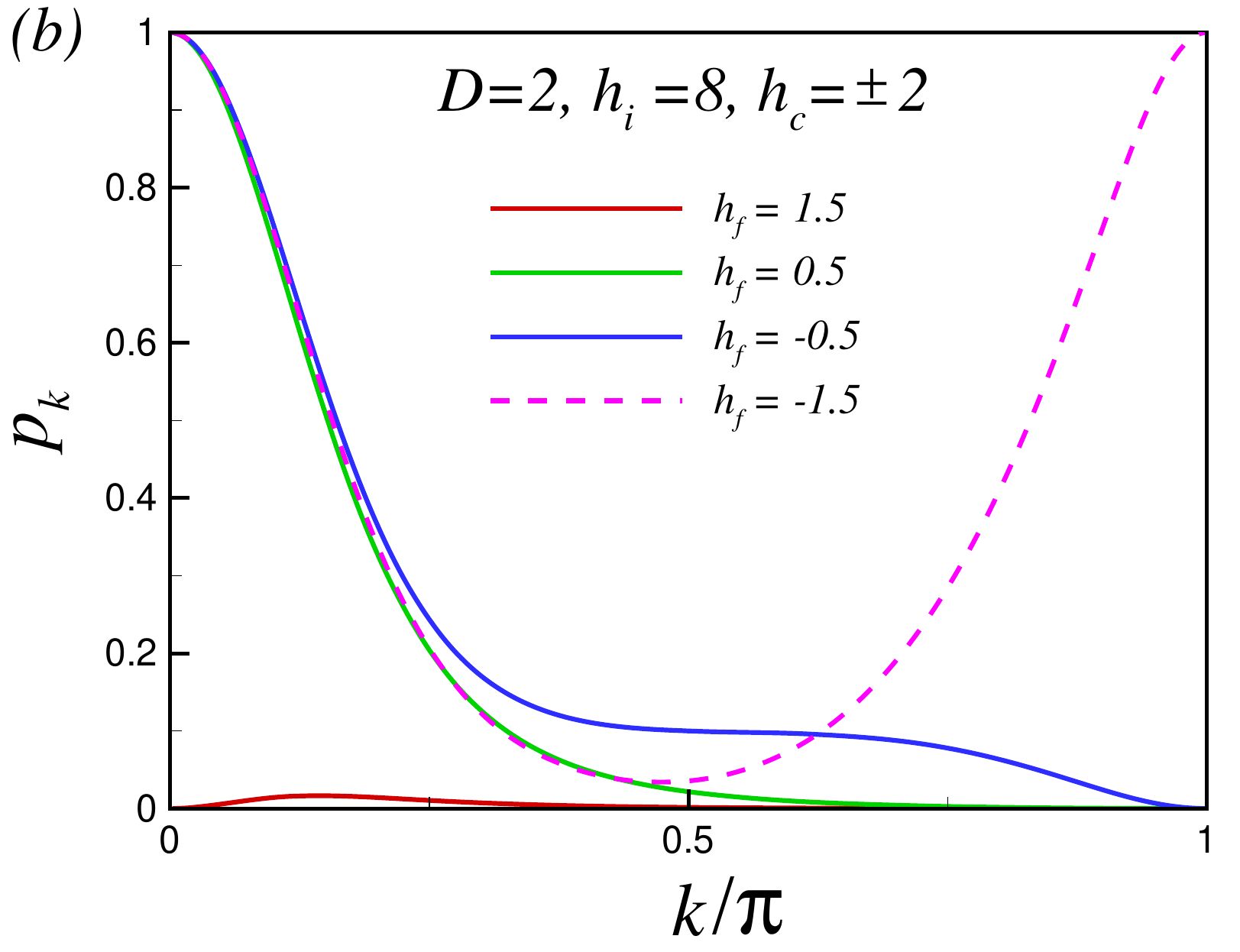}
\includegraphics[width=0.33\linewidth]{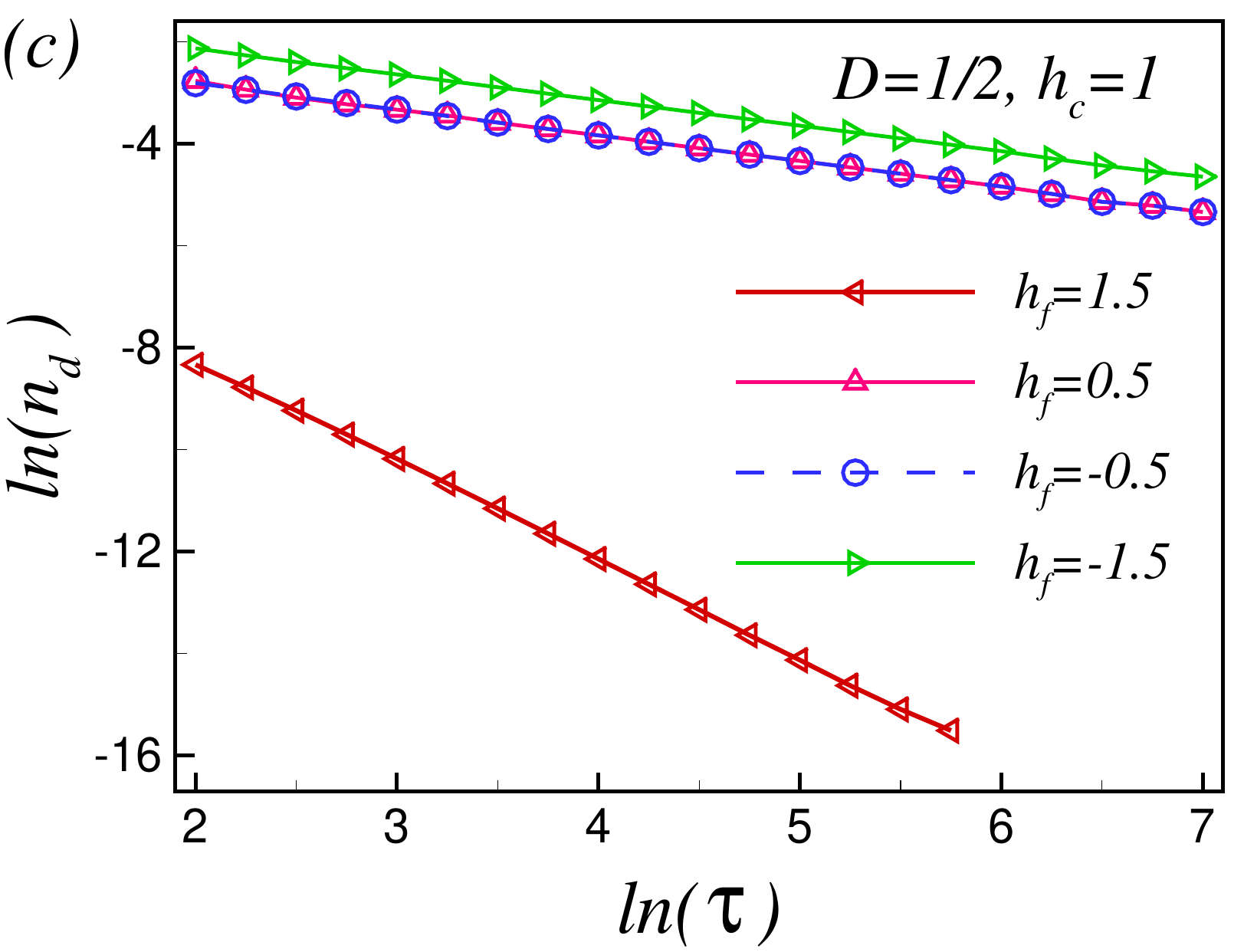}
}
\caption{(Color online) The transition probability of excitation to the upper level
in the transverse field Ising model with Dzyaloshinsky-Moriya (DM) interaction for
a quench from $h_i=8$ to the different values of quench end $h_f=1.5, 0.5, -0,5$ and $-1.5$,
for (a) $D=0.5$, and (b) $D=2$. (c) The transition probability of excitation to the upper level
in the generalized $XY$ model for different values of quench end $h_s^f=-0.5, -0.1, 0, 0.5$ and $5$, 
for a quench starting at $h_s^i=-5$.}
\label{figSM1}
\end{figure*}
%%%%%%%%%%%%%%%%%%%%%%%%%%%%%%%%%%%%%%%%%%%%%%%%%%%%%%%%%%%%%%%
%

\section*{C. The generalized XY model (GXY)}
The generalized $XY$ model dictated by the following Hamiltonian
%
%%%%%%%%%%%%%%%%%%%%%%%%%%%%%%%%%%%  Eq. S16 %%%%%%%%%%%%%%%%%%%%%%%%%%%%%%
\begin{eqnarray}
\label{eqS16}
{\cal H}^{GXY} = \!-\sum_{n=1}^N\Big[&&J (s_n^x s^x_{n+1} \!+\! s_n^y s^y_{n+1}) + (-1)^n h_{s} s^z_n 
- J_2  (s_n^x s^x_{n+2} + s_n^y s^y_{n+2})\,s^z_{s+1}\Big].
\end{eqnarray}
%%%%%%%%%%%%%%%%%%%%%%%%%%%%%%%%%%%%%%%%%%%%%%%%%%%%%%%%%%%%%%%%%%%%%%%%%%%
%
where, $N$ is the system size, $h_{s}$ represents the staggered transverse field, $J$ and $J_{3}$ are
exchange couplings between the spins on the nearest-neighbor and the next-nearest-neighbor sites respectively.

Performing the Jordan-Wigner fermionization and introducing the Nambu spinor $\Gamma^{\dagger}=(c^{q\dagger}_{k},c^{p\dagger}_{k})$,
the Fourier transformed Hamiltonian can be expressed in
Bogoliubov-de Gennes (BdG) form \cite{Titvinidze2003SM}, $H= \sum_{k\ge0}\Gamma^{\dagger}H(k)\Gamma$, with
%
%%%%%%%%%%%%%%%%%%%%%%%%%%%%%%%%%%%  Eq. S17 %%%%%%%%%%%%%%%%%%%%%%%%%%%%%%
\begin{eqnarray}
\label{eqS17}
H(k)=
\left(
  \begin{array}{cc}
    h_{s} & -J\cos(k/2)  \\
    \\
    -J\cos(k/2) &  -h_{s} \\
  \end{array}
\right)+\frac{J_{2}}{2}\cos(k)\mathbb{1},
\end{eqnarray}
%%%%%%%%%%%%%%%%%%%%%%%%%%%%%%%%%%%%%%%%%%%%%%%%%%%%%%%%%%%%%%%%%%%%%%
%
where $k=(2m-1)\pi/N$ with $-N/4+1\leqslant n\leqslant N/4$ for periodic boundary conditions \cite{Titvinidze2003SM}.
Using the standard Bogoliubov transformation
%
%%%%%%%%%%%%%%%%%%%%%%%%%%%%%%%%%%%%%%%%%%%%%%%%%%%%%%%%%%%%%%%%%%%%%%
\begin{eqnarray}
\nonumber
c_{k}^{q}=\cos(\frac{\theta_{k}(h_{s})}{2}) \alpha_{k}+\sin(\frac{\theta_{k}(h_{s})}{2}) \beta_{k},~~
c_{k}^{p}=-\sin(\frac{\theta_{k}(h_{s})}{2}) \alpha_{k}+ \cos(\frac{\theta_{k}(h_{s})}{2}) \beta_{k},
~~ \text{with} ~~\tan(\theta_{k}(h_{s}))\!=\!-J\cos(k/2)/h_{s},
\end{eqnarray}
%%%%%%%%%%%%%%%%%%%%%%%%%%%%%%%%%%%%%%%%%%%%%%%%%%%%%%%%%%%%%%%%%%%%%%
%
we finally can write the Hamiltonian in the diagonalized form as
${\cal H}\!=\!\sum_{k}[\varepsilon^{\alpha}_{k}(h_{s})\alpha^{\dagger}_{k} \alpha_{k}
+\varepsilon^{\beta}_{k}(h_{s})\beta^{\dagger}_{k}\beta_{k}]$,
where
%
%%%%%%%%%%%%%%%%%%%%%%%%%%%%%%%%%%%%%%%%%%%%%%%%%%%%%%%%%%%%%%%%%%%%%%
\begin{eqnarray}
\nonumber
e^{-}_{k}=\varepsilon^{\alpha}_{k}(h_{s})\!=\!(J_{2}/2)\cos(k)-\sqrt{(h_{s})^{2}+J^{2}\cos^{2}(k/2)},~~
e^{+}_{k}=\varepsilon^{\beta}_{k}(h_{s})\!=\!(J_{2}/2)\cos(k)+\sqrt{(h_{s})^{2}+J^{2}\cos^{2}(k/2)}.
\end{eqnarray}
%%%%%%%%%%%%%%%%%%%%%%%%%%%%%%%%%%%%%%%%%%%%%%%%%%%%%%%%%%%%%%%%%%%%%%
%
Searching the spectrum of the model reveals that the topology of the Fermi surface of the system changes at $h^{(c)}_s=\pm J_2/2$ which corresponds to a quantum phase transition in the ground state of the system. This transition separates an antiferromagnetic phase from a type-I spin-liquid phase (Fig. \ref{figSM2}(a)) \cite{Titvinidze2003SM}.
\\ 
%
%%%%%%%%%%%%%%%%%%%%%%%%%%%%  Fig.SM2 %%%%%%%%%%%%%%%%%%%%%%%%%%%%%%%%
\begin{figure*}[t]
\centerline{
\includegraphics[width=0.25\linewidth]{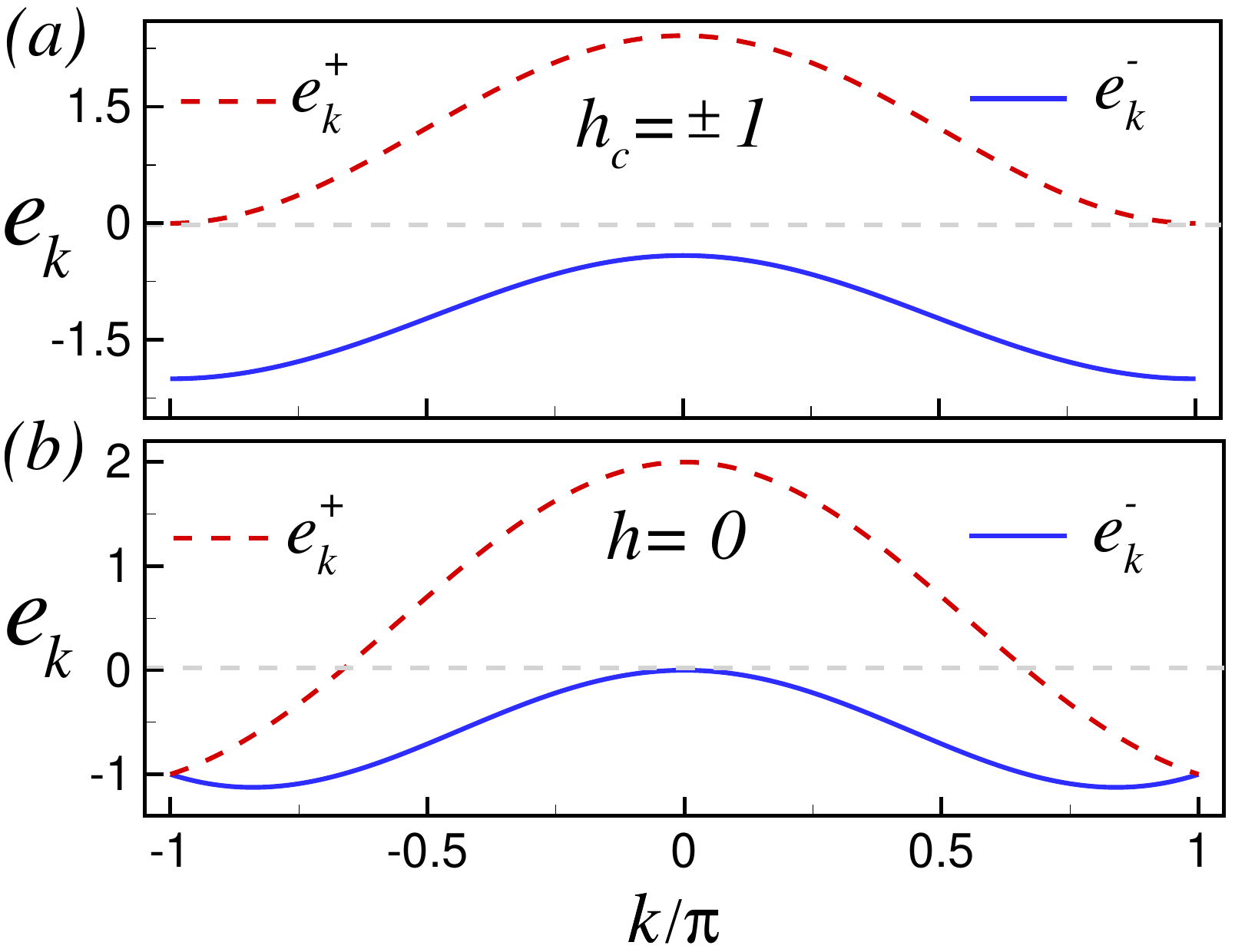}
\includegraphics[width=0.25\linewidth]{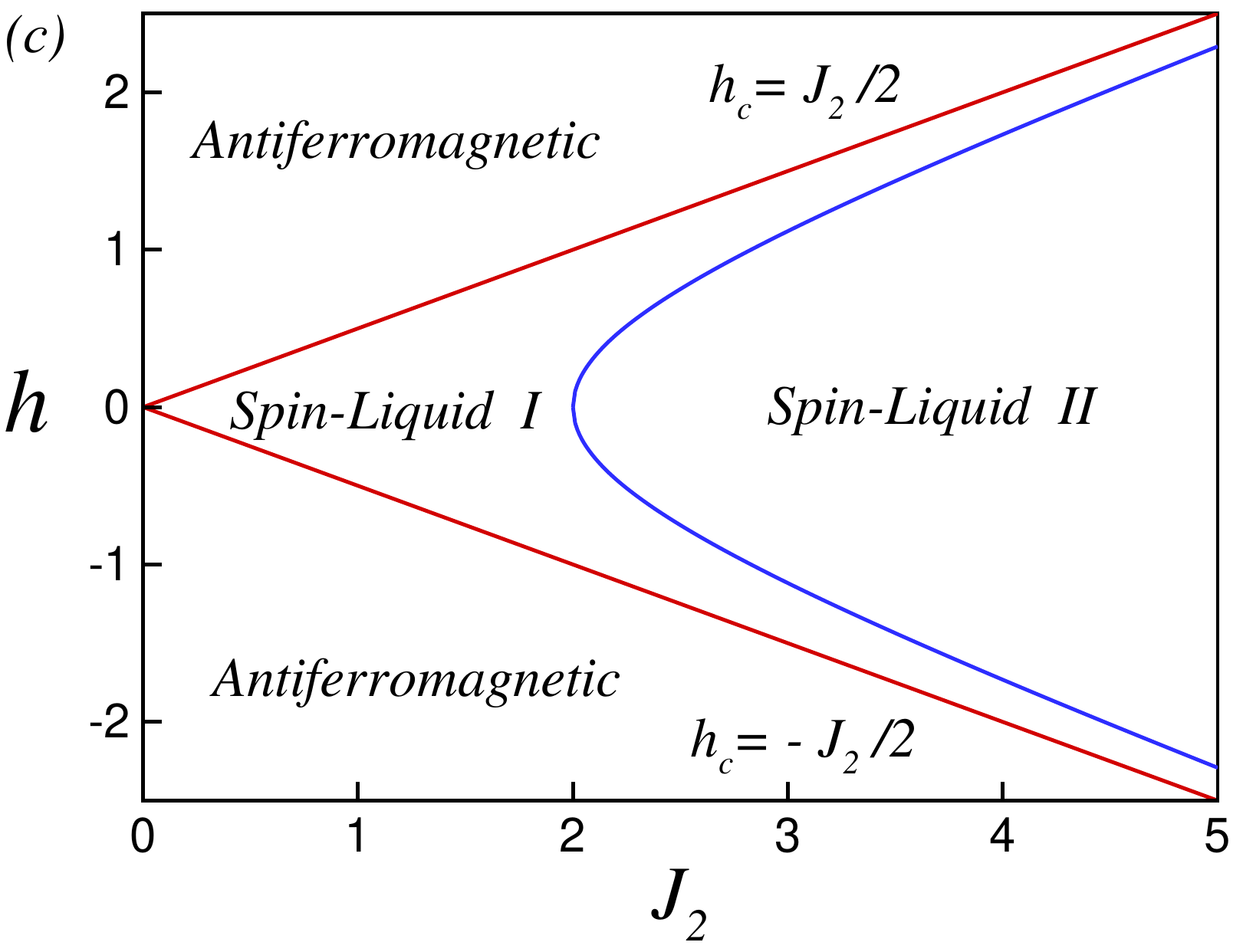}
\includegraphics[width=0.25\linewidth]{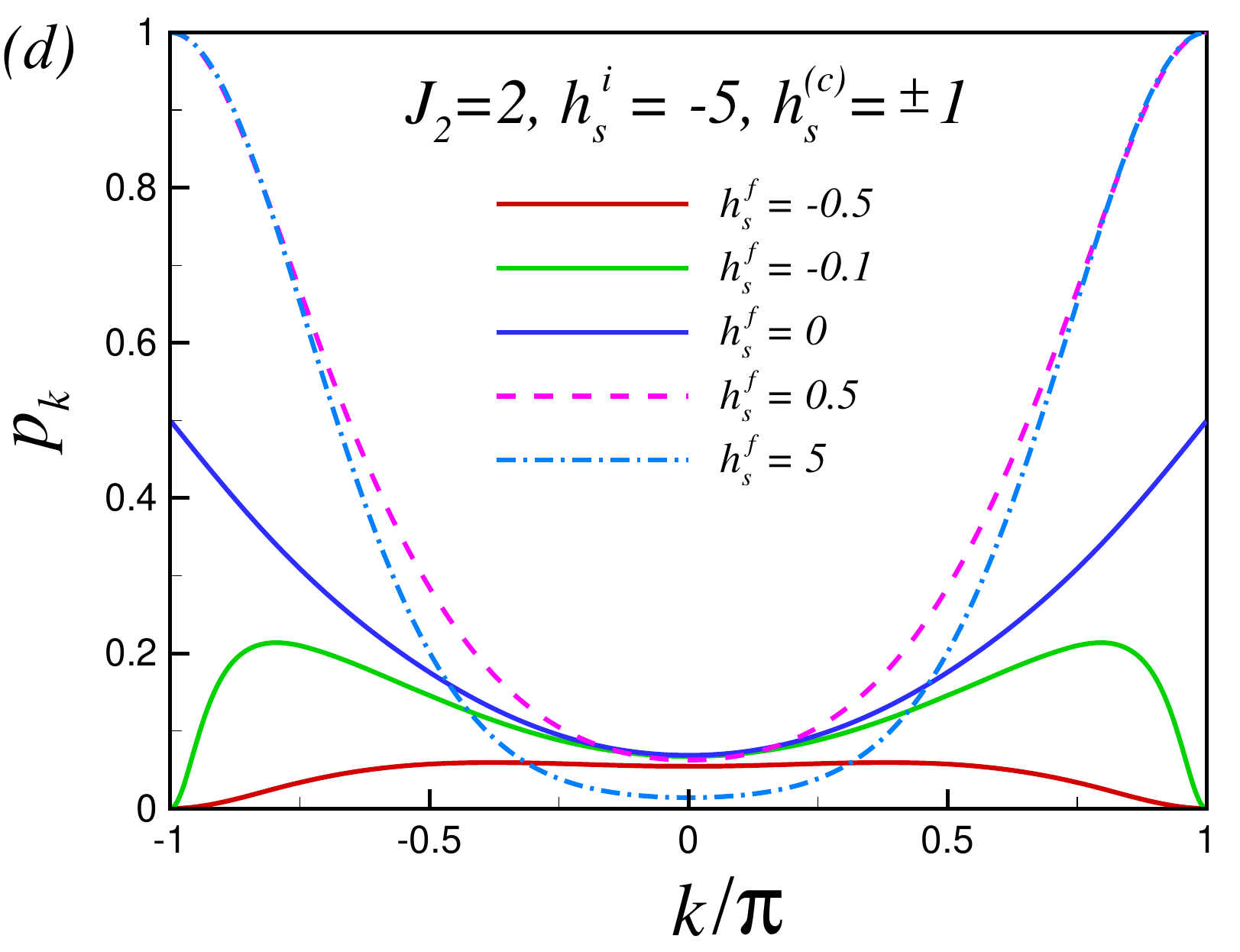}
\includegraphics[width=0.25\linewidth]{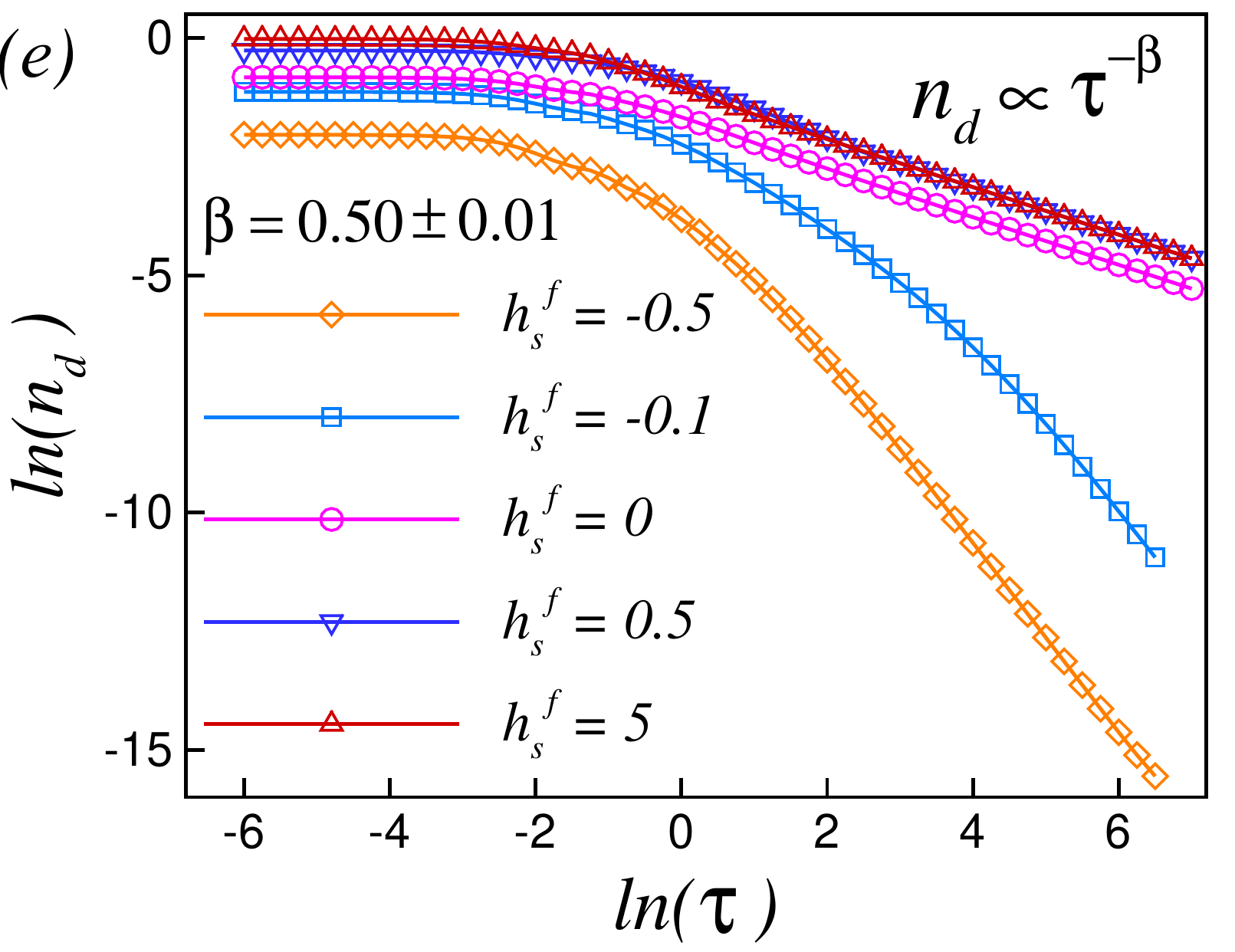}}
\caption{(Color online) (a) The phase diagram of generalized $XY$ model. (b) The transition probability of excitation to the upper level
in the generalized $XY$ model for different values of quench end $h_s^f=-0.5, -0.1, 0, 0.5$ and $5$, 
for a quench starting at $h_s^i=-5$. Panel (c) depicts defects density as a function of 
ramp time scale $\tau$ in generalized $XY$ model for $J_2=2$ for a quench from $h^i_s=-5$ 
to the different values of quench field end $h_s^f=-0.5, -0.1, 0, 0.5$ and $5$ for the system size $N=1024$. 
The quasi-particle spectrum $\pm e_{k}^{\pm}$ versus $k$ (d) at the critical point 
$h_c=\pm1$ and (e) at the non-critical point $h=0$.}
\label{figSM2}
\end{figure*}
%%%%%%%%%%%%%%%%%%%%%%%%%%%%%%%%%%%%%%%%%%%%%%%%%%%%%%%%%%%%%%%
%

For linear time dependent staggered filed $h_{s}(t)=t/\tau$ ($ t \in [h_s^i\tau, h_s^f\tau]$), the Hamiltonian in Eq. (\ref{eqS17}) for each mode has the form
${\cal H}_{k}^{GXY}(t)=t/\tau \sigma^x+\delta_k \sigma^z$, with $\delta_k=-J \cos(k/2)$, so transition rates can be calculated
analytically by the Landau-Zener formula \cite{ZenerSM,LandauSM,Vitanov1999SM}.
The density of defects can also be calculated numerically as $n_d=1-\frac{1}{N}\sum_k\langle\psi_{g,k}(t_f)|\rho_k(t_f)|\psi_{g,k}(t_f)\rangle$
where $|\psi_{g,k}(t_f)\rangle$ is the ground state of the Hamiltonian at $t_f$, and the density matrix $\rho(t)$ is calculated numerically using the Von-Neumann equation $i {\dot{\rho}_k}(t)=[{\cal H}_k^{GXY}(t),\rho_k(t)]$.
In Fig. \ref{figSM2}(b) the transition probability has been shown versus $k$ for a quench from $h_i=-5$ to the different values of quench end field $h_f=-0.5, -0.1, 0, 0.5$ and $5$ for $\tau=1$. Despite our expectation the transition for a quench across the critical point is small while it is significant for a quench across the non-critical point $h=0$ at modes $k=0, \pi$ where the quasi-particle band becomes gapless.

The numerical simulation of defect density has been plotted in Fig. \ref{figSM2}(c) for $J_2=2$ ($h_s^{(c)}=\pm1$) 
for different values of quench filed end $h_{s}^f$ with the quench starting at $h_s^i=-5$.
It can be seen that, when quench crosses the critical point $h^{(c)}_s=-1$ ($h_{s}^f=-0.5, -0.1$) although the density of defects decreases by increasing the ramp time scales, this trend deviates from KZ scaling and exhibits a much accelerated dependence on the ramp time scale. While for quench to/throught the {\em non-critical} point $h_{s}=0$ ($h_{s}^f=0, 0.5, 5$), numerical simulations verify the KZ scaling law $n_d\propto \tau^{-\beta}$ with $\beta=0.5\pm0.01$. Our analysis of the GCM and the TFIMDM lead us to address this challenge by probing the quasi-particle spectrum of the model (Fig. \ref{figSM2}(d)-(e)). As a quench crosses the QCP, the accelerated decay of defect density, surpassing the KZ prediction, can be attributed to the gapped quasi-particle states that dictate the dynamics. As one approaches the QCP, these quasi-particles remain completely gapped, resulting in adiabatic driving (Fig. \ref{figSM2}(d)).
In contrast, the correspondence between defect density scaling and the KZM prediction for a quench through a {\em non-critical} point can be explained by the massless quasi-particles that appear at a {\em non-critical} point (Fig. \ref{figSM2}(e)).

%\bibliography{AKZ_References}

%

\end{widetext}

\end{document}